\documentclass[12pt]{article}
\usepackage{color}
\usepackage{graphicx}
\def\nabstar#1{\nabla\kern-0.5pt\smash{\raise 4.5pt\hbox{$\ast$}}
               \kern-4.5pt_{#1}}

\def\drvstar#1{\partial\kern-0.5pt\smash{\raise 4.5pt\hbox{$\ast$}}
               \kern-5.0pt_{#1}}

\def\newline{\relax\ifhmode\null\hfil\break\else\nonhmodeerr@\newline\fi}
\def\frac#1#2{{#1\over#2}}
\def\text#1{{\hbox{\rm #1}}}
\def\flushpar{{\par \noindent}}

\newcommand{\beq}{\begin{equation}}
\newcommand{\eeq}{\end{equation}}
\newcommand{\bea}{\begin{eqnarray}}
\newcommand{\eea}{\end{eqnarray}}

\def\EQ{\hspace{-2mm} &=& \hspace{-2mm}}

\def\BA{\begin{eqnarray}}
\def\EA{\end{eqnarray}}
\def\BAN{\begin{eqnarray*}}
\def\EAN{\end{eqnarray*}}

\def\nn{\nonumber\\}

\def\g5{\gamma_5}
\def\g4{\gamma_4}
\def\g3{\gamma_3}
\def\g2{\gamma_2}
\def\g1{\gamma_1}
\def\gi{\gamma_i}

\def\u{{\bf u}}

\def\s{{\bf s}}
\def\c{{\bf c}}
\def\q{{\bf q}}
\def\Q{{\bf Q}}
\def\ubar{\bar{\bf u}}

\def\sbar{\bar{\bf s}}
\def\cbar{\bar{\bf c}}
\def\qbar{\bar{\bf q}}

\input epsf.sty
\newdimen\psfigsize
\def\psfigure#1 #2 #3 #4 #5{
    \begin{figure}[tbh]
      \begin{center}
      \vbox{
        \null\vskip-0.2in\hskip#2
        \epsfxsize=#1
        \epsfbox{#4}
        \vskip -0.3in
        \caption {#5 \label{#3}}
        \vskip 0.0 true in plus 0.3 true in
      }
      \end{center}
   \end{figure}
}
\usepackage{graphicx}
\begin{document}
\thispagestyle{empty}
\begin{flushright}
NTUTH-05-505F \\
December 2005 \\
\end{flushright}
\vskip 2.5truecm
\centerline{{\LARGE $ Y(4260) $ on the lattice}}
\vskip 1.0truecm
\centerline{{\bf Ting-Wai~Chiu$^{1}$, Tung-Han~Hsieh$^{2}$}}
\vskip2.0ex
\centerline{$^1\hskip-3pt$ \it
Department of Physics, National Taiwan University,}
\vskip1.0ex
\centerline{\it Taipei, 10617, Taiwan}
\vskip2.0ex
\centerline{$^2\hskip-3pt$ \it
Physics Section, Commission of General Education,}
\vskip1.0ex
\centerline{\it National United University, Miao-Li, 36003, Taiwan}
\vskip1.0ex
\vskip2.0ex
\centerline{\bf (TWQCD Collaboration)}
\vskip 1cm
\bigskip \nopagebreak \begin{abstract}

\noindent

We investigate the mass spectra of closed-charm mesons with 
$ J^{PC} = 1^{--} $, for hybrid charmonium, 
molecules, and diquark-antidiquark operators,  
in quenched lattice QCD with exact chiral symmetry.
For two lattice volumes $ 24^3 \times 48 $ and $ 20^3 \times 40 $, 
each of 100 gauge configurations generated with single-plaquette action 
at $ \beta = 6.1 $, we compute point-to-point quark propagators   
and measure the time-correlation functions of 
these exotic meson operators. For the molecular operator 
$ \{ (\qbar\gamma_5\gamma_i\c)(\cbar\gamma_5\q)-
     (\cbar\gamma_5\gamma_i\q)(\qbar\gamma_5\c) \} $, 
it detects a resonance with mass around $ 4238 \pm 31 $ MeV, 
which is naturally identified with $ Y(4260) $. 
Further, for any molecular and diquark-antidiquark operator, 
it detects heavier exotic charmed mesons,  
with $ (\c\s\cbar\sbar) $ around $ 4450 \pm 100 $ MeV, and  
$ (\c\c\cbar\cbar) $ around $ 6400 \pm 50 $ MeV.


Charmed Mesons, Diquarks

\end{abstract}
\vskip 1.5cm 
\newpage\setcounter{page}1

\section{Introduction}

Recently, a new state $ Y(4260) $ with 
a mass of $ 4259(8)(4) $ MeV and a width $ 88(23)(5) $ MeV, 
in association with an initial state radiation photon, 
has been observed by BaBar collaboration 
in $ e^- e^+ $ annihilation \cite{Aubert:2005rm}.
This immediately implies that its $ J^{PC} = 1^{--} $. 
From the experimental and theoretical spectrum of ($ \c\cbar $) states, 
$ Y(4260) $ can hardly be interpreted as one of the
radial and/or orbital excitations of ($ \c\cbar $). 
Thus it is most likely an exotic (non-$q\bar q $) meson.

So far, theoretical models to understand $ Y(4260) $ are: 
(i) hybrid charmonium $ \cbar\c g $ \cite{Zhu:2005hp,Close:2005iz},
(ii) P-wave excitation of the diquark-antidiquark $ [\c\s][\cbar\sbar] $ 
\cite{Maiani:2005pe}; 
and (iii) molecule composed of 2 mesons, e.g., 
$ \rho \chi_{c1} $ \cite{Liu:2005ay}.

Now the important question is whether there is 
a resonance around 4260 MeV with $ J^{PC} = 1^{--} $,  
in the spectrum of QCD. 
In this paper, we investigate the mass spectra of several 
interpolating operators whose lowest-lying states have $ J^{PC} = 1^{--} $,  
in lattice QCD with exact chiral symmetry
\cite{Kaplan:1992bt,Narayanan:1995gw,Neuberger:1997fp,
      Ginsparg:1981bj,Chiu:2002ir}.
For two lattice volumes $ 24^3 \times 48 $ and $ 20^3 \times 40 $, 
each of 100 gauge configurations generated with single-plaquette action 
at $ \beta = 6.1 $, we compute point-to-point quark propagators   
for 30 quark masses in the range $ 0.03 \le m_q a \le 0.80 $, 
and measure the time-correlation functions of the exotic meson operators 
which can overlap with $ Y(4260) $.  
The inverse lattice spacing $ a^{-1} $
is determined with the experimental input of $ f_\pi $,
while the strange quark bare mass $ m_s a = 0.08 $, and the charm
quark bare mass $ m_c a = 0.80 $ are fixed such that the masses of
the corresponding vector mesons are in good agreement with  
$ \phi(1020) $ and $ J/\psi(3097) $ respectively 
\cite{Chiu:2005zc}. Our scheme of computing quark propagators 
has been outlined in \cite{Chiu:2003iw}.

Note that we are working in the quenched approximation which 
in principle is unphysical. However, our previous results on 
charmed baryon masses, and also charmed meson masses and decay constants 
(theoretical predictions) \cite{Chiu:2005zc} turn out to be in good 
agreement with the experimental values. This seems to suggest that it is
plausible to use the quenched lattice QCD with exact chiral symmetry
to investigate the mass spectra of the charmed meson operators 
constructed in this paper, as a first step toward the unquenched calculations.
The systematic error due to quenching can only be determined
after we can repeat the same calculation with unquenched
gauge configurations. However, the Monte Carlo simulation
of unquenched gauge configurations for lattice QCD with exact
chiral symmetry, on the lattices $ 20^3 \times 40 $ and $ 24^3 \times 48 $
at $ \beta = 6.1 $, still remains a great challenge to the lattice community. 
Thus, in this paper, we proceed with the quenched approximation,
assuming that the quenching error does not change our conclusions
dramatically, in view of the good agreement between our previous
quenched mass spectra of charmed hadrons \cite{Chiu:2005zc}
and the experimental values.


\section{The Hybrid Charmonium $ \cbar \c g $}

The local interpolating operators for
hybrid charmonium with $ J^{PC} = 1^{--} $ can be constructed
as $ \cbar F_{4i} \c $, $ \epsilon_{ijk} \cbar \gamma_5 F_{jk} \c $, and
$ \epsilon_{ijk} \cbar \gamma_4 \gamma_5 F_{jk} \c $.
Here the matrix-valued gluon field tensor $ F_{\mu\nu}(x) $
can be obtained from the four plaquettes surrounding $ x $ on the
($ \hat\mu, \hat\nu $) plane, i.e.,
\BAN
g a^2 F_{\mu\nu}(x)
\hspace{-2mm}
& \simeq & \hspace{-2mm}
\frac{1}{8i}
   [  P_{\mu\nu}(x) + P_{\mu\nu}(x-\hat\mu) + P_{\mu\nu}(x-\hat\nu)
      + P_{\mu\nu}(x-\hat\mu-\hat\nu)  \nn
& &
      - P^{\dagger}_{\mu\nu}(x) - P^{\dagger}_{\mu\nu}(x-\hat\mu)
      - P^{\dagger}_{\mu\nu}(x-\hat\nu)
      - P^{\dagger}_{\mu\nu}(x-\hat\mu-\hat\nu) ]
\EAN
where
\BAN
P_{\mu\nu}(x) = U_\mu(x) U_\nu(x+\hat\mu)
                U^{\dagger}_\mu(x+\hat\nu) U^{\dagger}_\nu(x) \ .
\EAN

\begin{figure}[htb]
\begin{center}
\begin{tabular}{@{}cc@{}}
\includegraphics*[height=9cm,width=7cm]{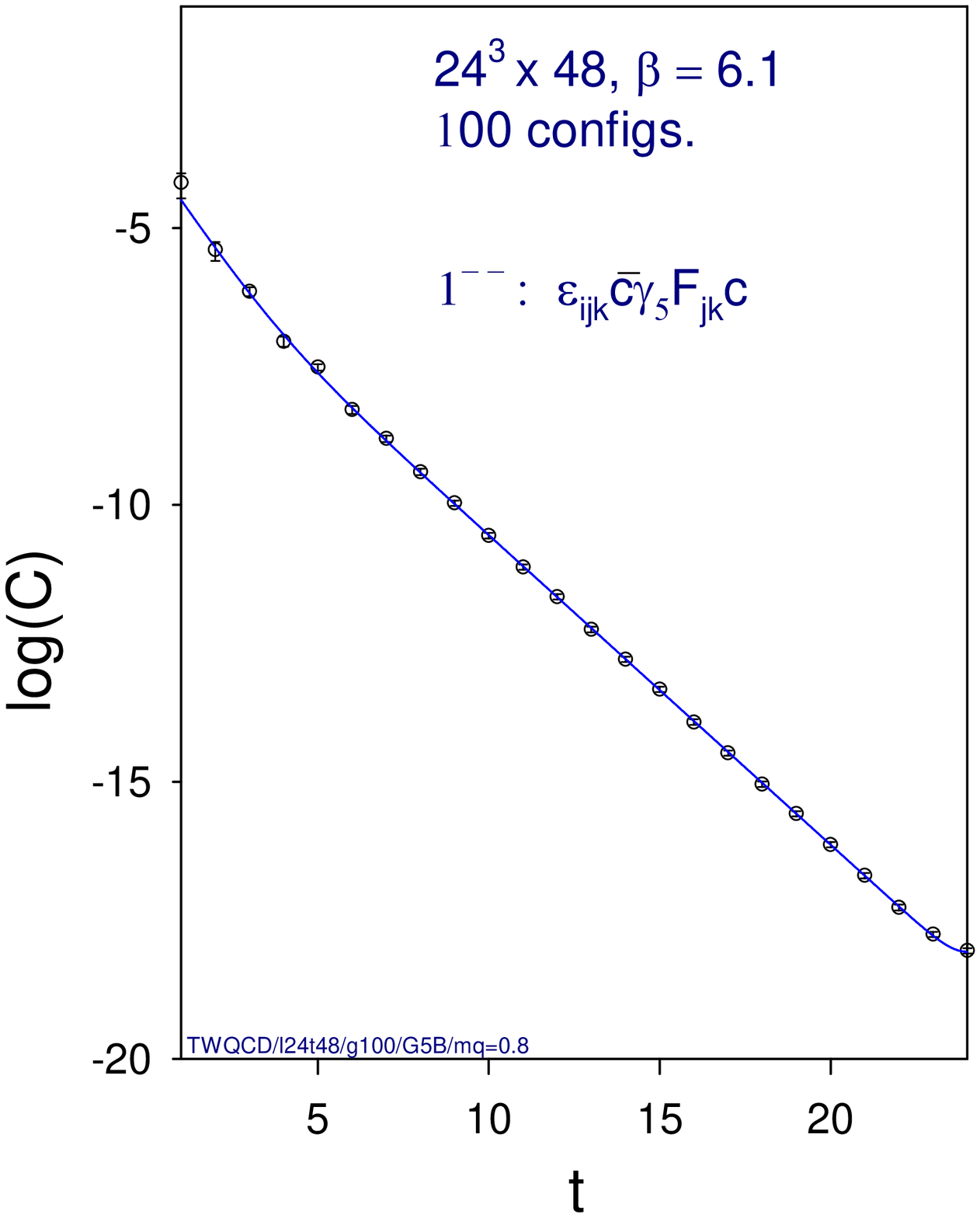}
&
\includegraphics*[height=9cm,width=7cm]{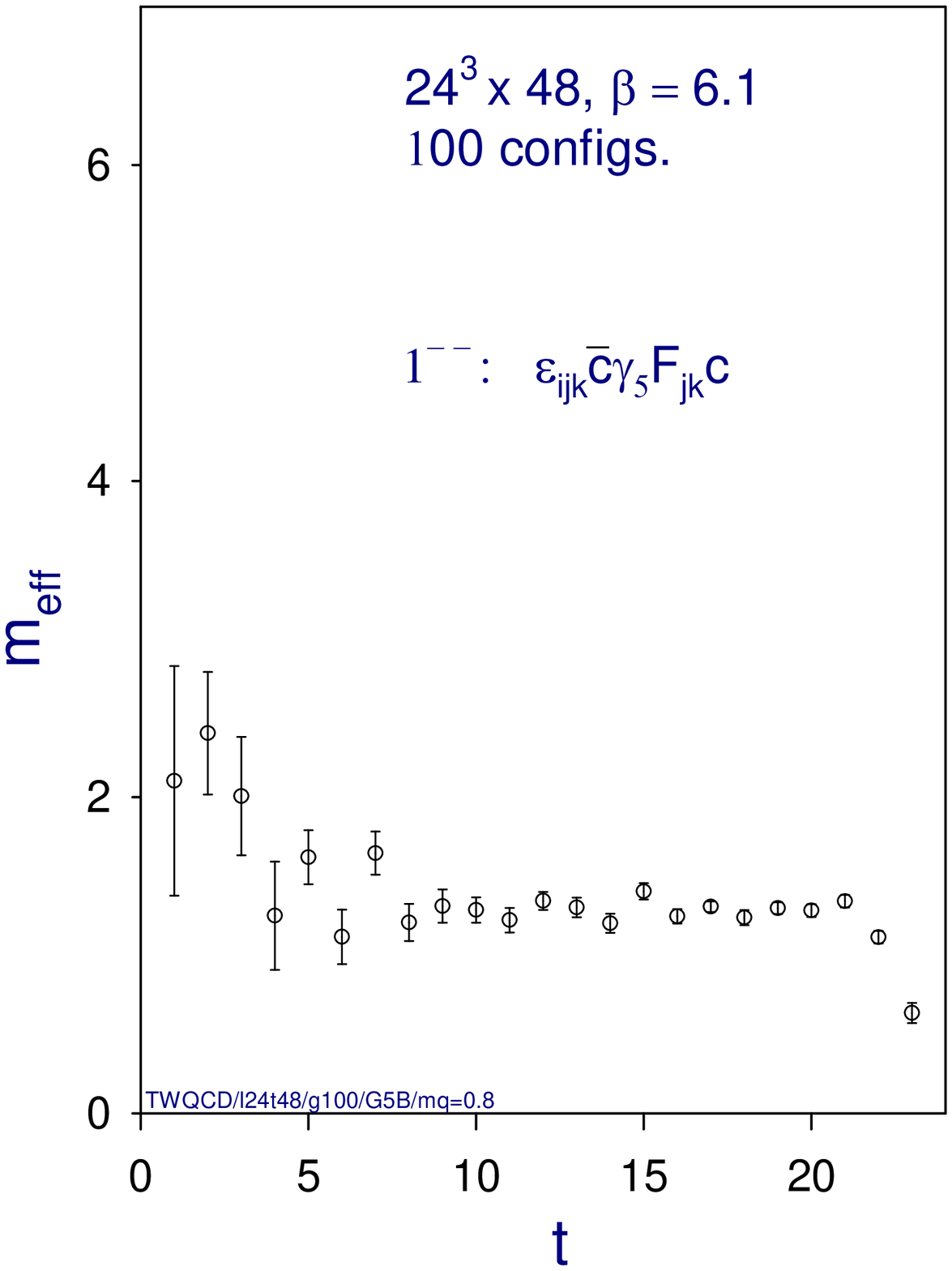}
\\ (a) & (b)
\end{tabular}
\caption{
(a) The time-correlation function $ C(t) $ of the 
hybrid meson operator $ \epsilon_{ijk} \cbar \gamma_5 F_{jk} \c $ with
$ J^{PC} = 1^{--} $, on the $ 24^3 \times 48 $ lattice at $ \beta = 6.1 $.
The solid line is the double hyperbolic-cosine fit for $ t \in [1,24] $.
(b) The effective mass $ M_{eff}(t) = \ln [C(t)/C(t+1)]  $
of $ C(t) $ in Fig.\ \ref{fig:G5B_080}a.
}
\label{fig:G5B_080}
\end{center}
\end{figure}

In Fig. \ref{fig:G5B_080}, the time-correlation function $ C(t) $
of the hybrid charmonium operator
$ \epsilon_{ijk} \cbar \gamma_5 F_{jk} \c $ is plotted, 
together with the effective mass $ \ln [C(t)/C(t+1)] $, 
for 100 gauge configurations generated with single-plaquette action 
on $ 24^3 \times 48 $ at $ \beta = 6.1 $. 
Here $ C(t) $ has been averaged over $ i=1,2,3 $, where in each case, 
the ``forward-propagator" $ C_i(t) $ and ``backward-propagator" 
$ C_i(T-t) $ are averaged to increase the statistics. 
The same strategy is applied to all time-correlation functions
in this paper. The solid line in (a) is the double hyperbolic-cosine fit  
\BAN
W_1 \left( e^{-m_1 a t} + e^{-m_1 a (T-t)} \right) +
W_2 \left( e^{-m_2 a t} + e^{-m_2 a (T-t)} \right)
\EAN
for $ t \in [1,24] $. 
It gives $ m_1 = 2977(28) $ MeV and $ m_2 = 4501(178) $ MeV
with $ \chi^2/d.o.f. = 0.95 $. In this paper, we have adopted 
the procedure outlined in Ref. \cite{Iwasaki:1995cm}
to perform our data fitting and error estimation.   
Now we identity the lowest-lying state with 
$ J/\psi $, since the hybrid charmonium operator with $ 1^{--} $ also 
overlaps with $ J/\psi $. Then the first excited state with mass 
$ 4501(178) $ MeV is identified with the lowest-lying hybrid charmonium 
with $ 1^{--} $. Note that the error of the mass of the first excited state 
is relatively large, since it is obtained by double hyperbolic-cosine fit. 
Evidently the mass of the lowest-lying hybrid charmonium state with $ 1^{--} $ 
is higher than 4260 MeV. Thus it is unlikely to be identified with Y(4260), 
even though we could not rule out such a possibility, 
due to the large error bar.
Nevertheless, we will report a more precise determination of the mass 
of the hybrid charmonium with $ 1^{--} $ in a future publication.

\section{The Molecular Operators}

In this section, we construct three molecular operators with quark 
content ($ \c\q\cbar\qbar $) such that the lowest-lying state of 
each operator has $ J^{PC} = 1^{--} $. Then we compute the 
time correlation function of each operator, and extract the
mass of its lowest-lying state. Explicitly, these molcular operators are: 
\bea
\label{eq:DVDIS}
M_1 \EQ \frac{1}{\sqrt{2}} \left\{ (\qbar\gi\c)(\cbar\q)
                                  +(\cbar\gi\q)(\qbar\c) \right\} \\
\label{eq:JpiI}
M_2 \EQ (\cbar\gi\c)(\qbar\q) \\         
\label{eq:D5VDA}
M_3 \EQ \frac{1}{\sqrt{2}} \left\{
               (\qbar\gamma_5\gamma_i\c)(\cbar\gamma_5\q)
              -(\cbar\gamma_5\gamma_i\q)(\qbar\gamma_5\c) \right\} 
\eea
%
%
%
%

\begin{figure}[htb]
\begin{center}
\includegraphics*[height=9cm,width=7cm]{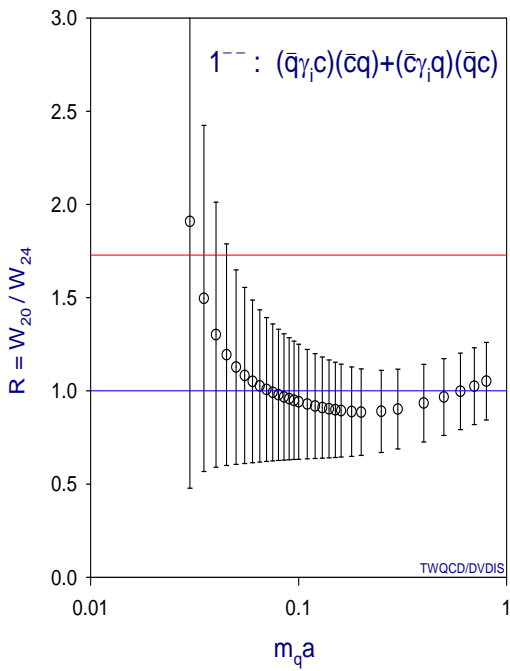}
\caption{
The ratio of spectral weights of the lowest-lying state
of the molecular operator $ M_1 $,
for $ 20^3 \times 40 $ and $ 24^3 \times 48 $ lattices at $ \beta = 6.1 $.
The upper-horizontal line $ R = (24/20)^3 = 1.728 $,
is the signature of 2-particle scattering state,
while the lower-horizontal line $ R = 1.0 $ is the signature
of a resonance.}
\label{fig:sw2024_DVDIS}
\end{center}
\end{figure}

\begin{figure}[htb]
\begin{center}
\includegraphics*[height=9cm,width=7cm]{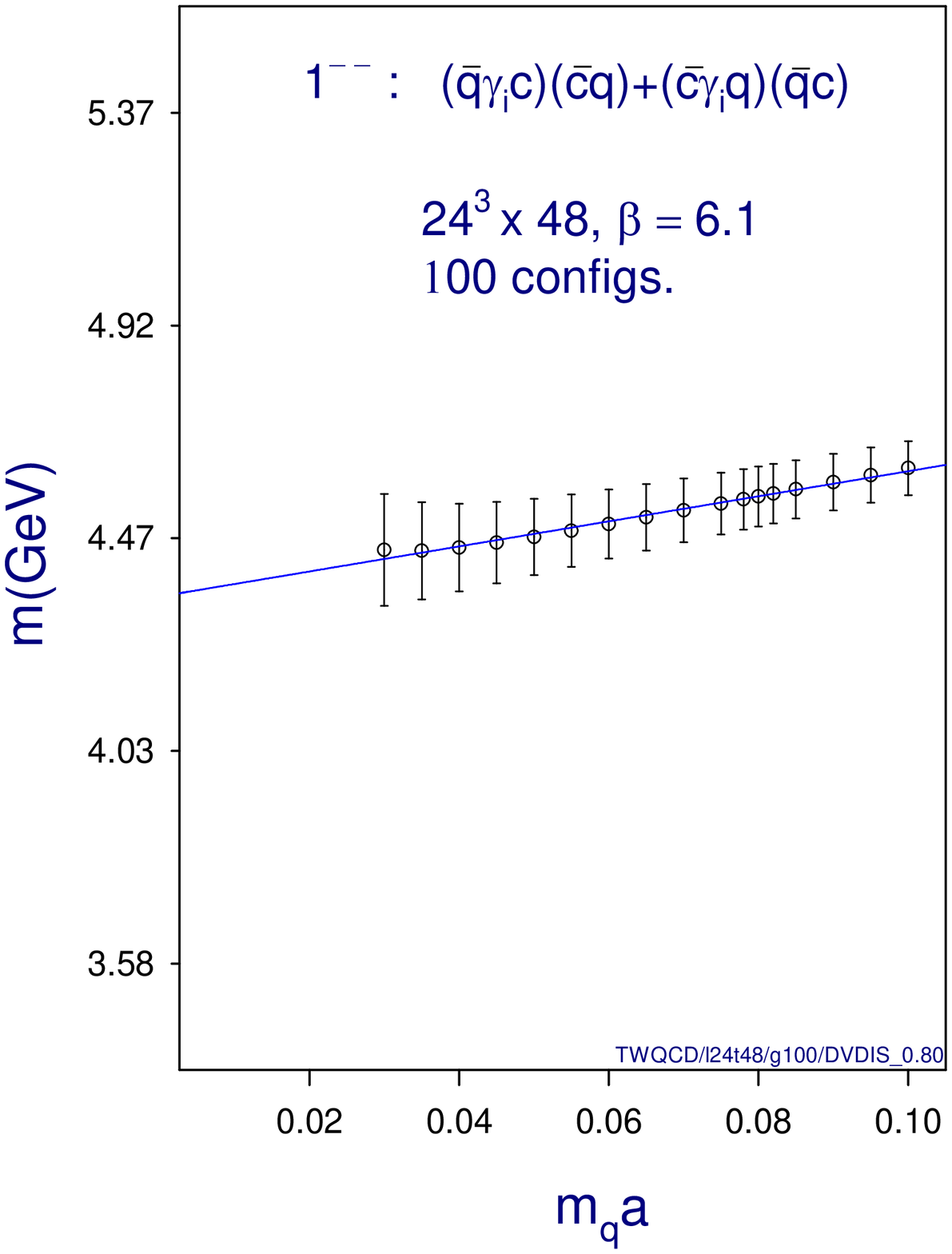}
\caption{
The mass of the lowest-lying state
of $ M_1 $ versus the quark mass $ m_q a $, 
on the $ 24^3 \times 48 $ lattice at $ \beta = 6.1 $.
The solid line is the linear fit with $ \chi^2 /d.o.f. = 0.34 $.}
\label{fig:mass_DVDIS}
\end{center}
\end{figure}

\begin{figure}[htb]
\begin{center}
\begin{tabular}{@{}cc@{}}
\includegraphics*[height=9cm,width=7cm]{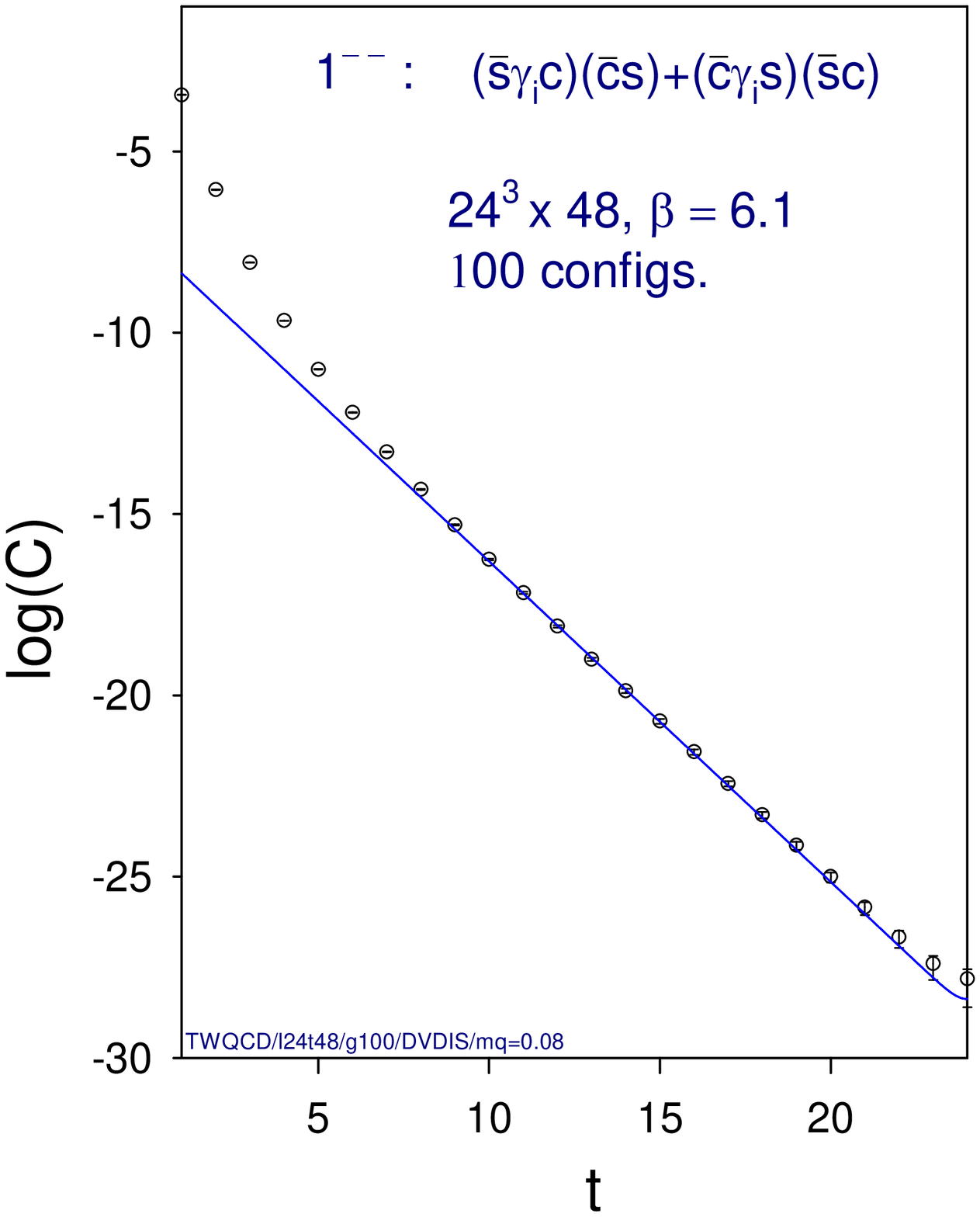}
&
\includegraphics*[height=9cm,width=7cm]{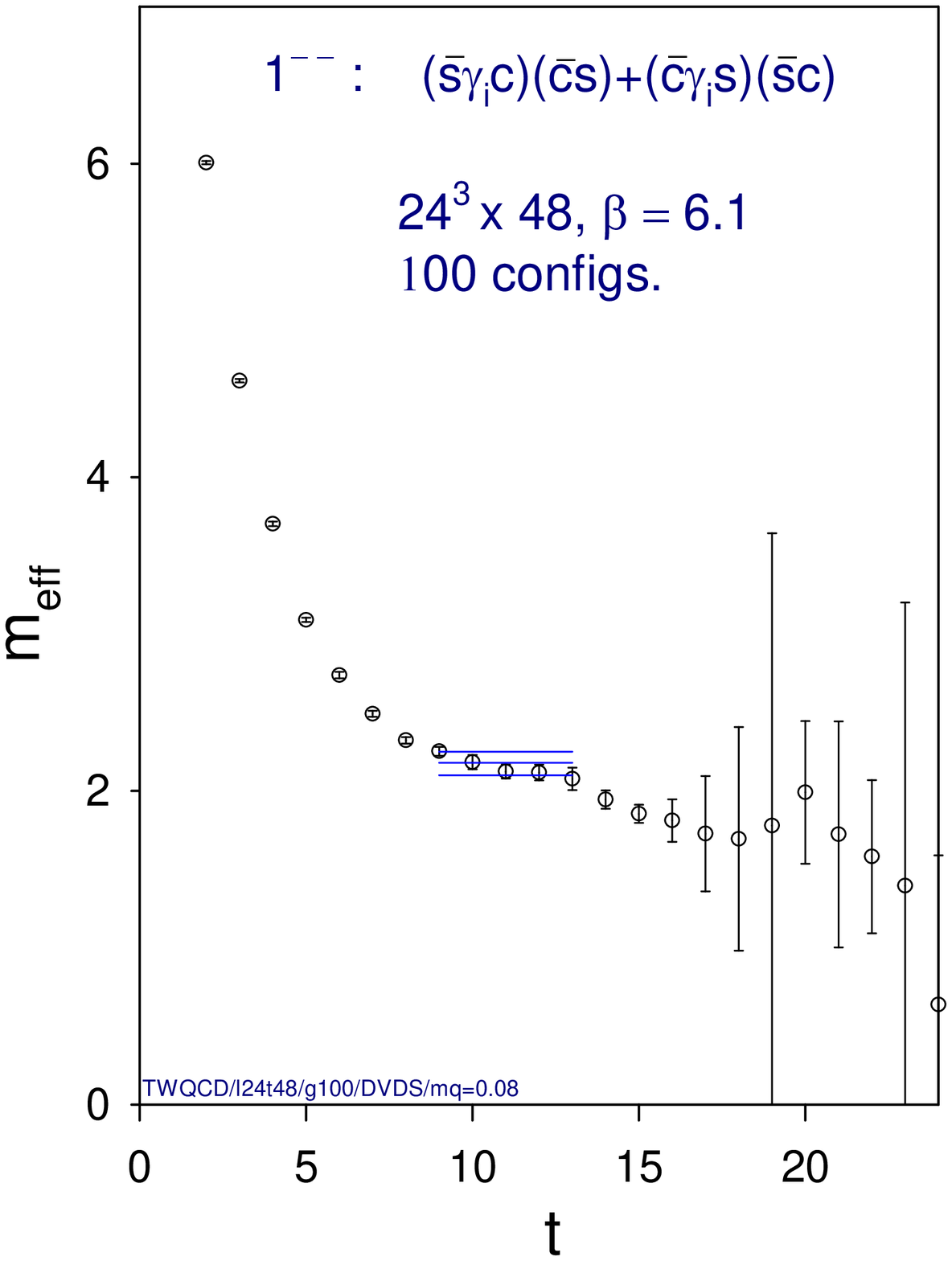}
\\ (a) & (b)
\end{tabular}
\caption{
(a) The time-correlation function $ C(t) $ of the lowest-lying state
of $ M_1 $ for $ m_q = m_s = 0.08 a^{-1} $,  
on the $ 24^3 \times 48 $ lattice at $ \beta = 6.1 $.
The solid line is the hyperbolic-cosine fit for $ t \in [9,13] $
with $ \chi^2/d.o.f. = 0.97 $.
(b) The effective mass $ M_{eff}(t) = \ln [C(t)/C(t+1)]  $
of $ C(t) $ in Fig.\ \ref{fig:DVDIS_008}a.
}
\label{fig:DVDIS_008}
\end{center}
\end{figure}

\begin{figure}[htb]
\begin{center}
\begin{tabular}{@{}cc@{}}
\includegraphics*[height=9cm,width=7cm]{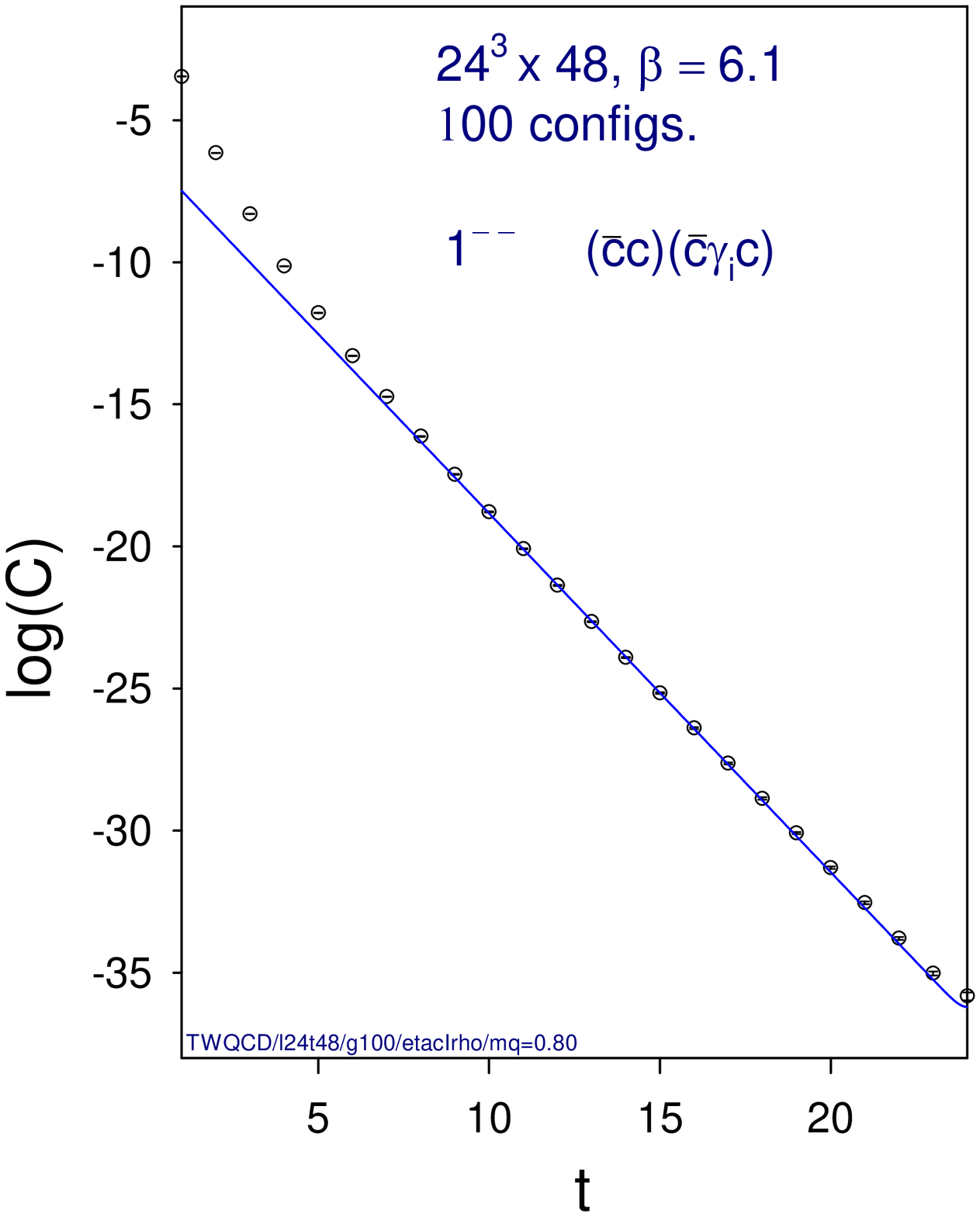}
&
\includegraphics*[height=9cm,width=7cm]{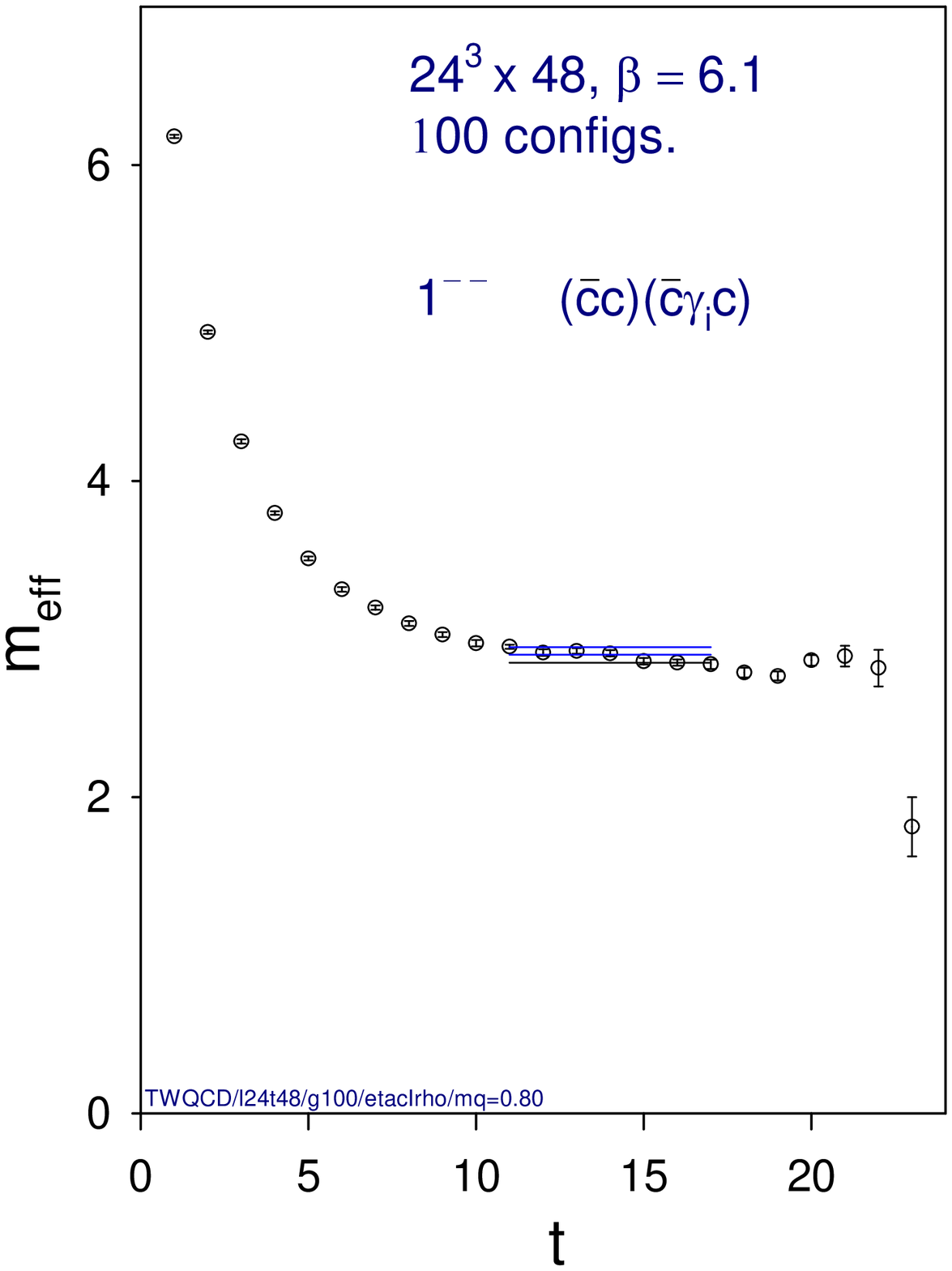}
\\ (a) & (b)
\end{tabular}
\caption{
(a) The time-correlation function $ C(t) $ of the lowest-lying state
of the molecular operator $ (\cbar\gi\c)(\cbar\c) $,    
on the $ 24^3 \times 48 $ lattice at $ \beta = 6.1 $.
The solid line is the hyperbolic-cosine fit for $ t \in [11,17] $
with $ \chi^2/d.o.f. = 0.61 $.
(b) The effective mass $ M_{eff}(t) = \ln [C(t)/C(t+1)]  $
of $ C(t) $ in Fig.\ \ref{fig:DVDIS_080}a.
}
\label{fig:DVDIS_080}
\end{center}
\end{figure}

\begin{figure}[htb]
\begin{center}
\begin{tabular}{@{}cc@{}}
\includegraphics*[height=9cm,width=7cm]{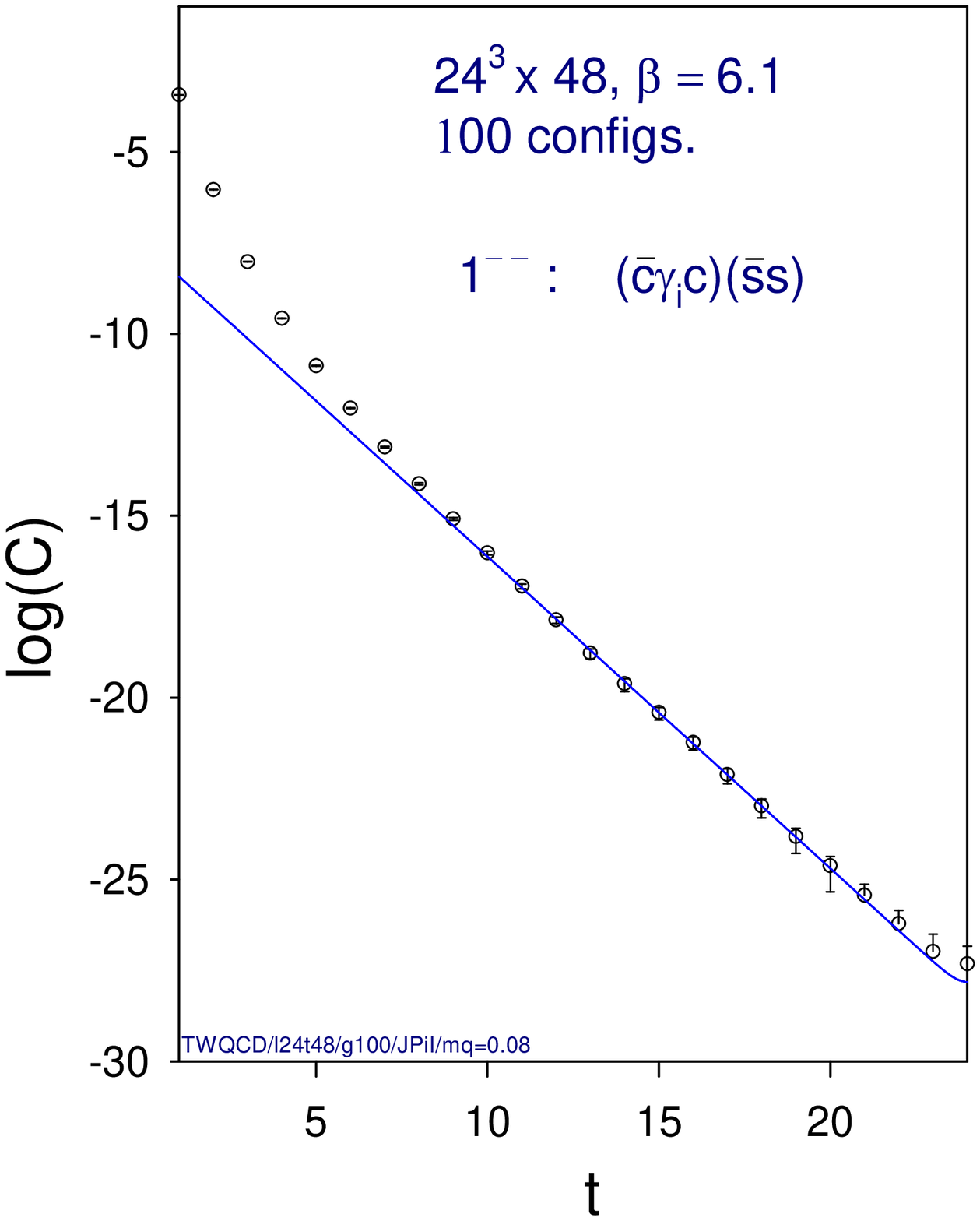}
&
\includegraphics*[height=9cm,width=7cm]{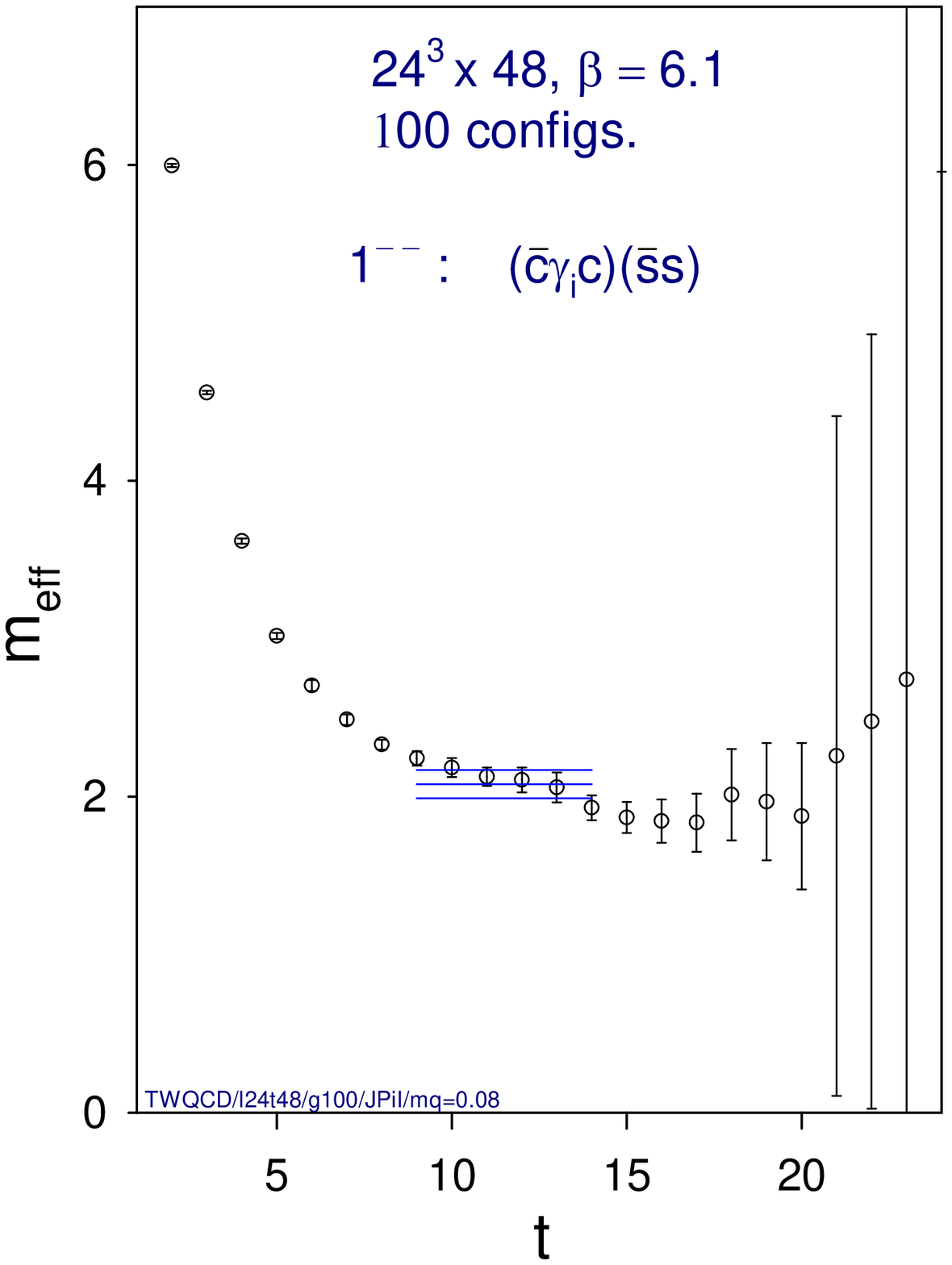}
\\ (a) & (b)
\end{tabular}
\caption{
(a) The time-correlation function $ C(t) $ of the lowest-lying state
of $ (\cbar \gi \c) (\sbar\s) $,
on the $ 24^3 \times 48 $ lattice at $ \beta = 6.1 $.
The solid line is the hyperbolic-cosine fit for $ t \in [9,14] $
with $ \chi^2/d.o.f. = 0.56 $.
(b) The effective mass $ M_{eff}(t) = \ln [C(t)/C(t+1)]  $
of $ C(t) $ in Fig.\ \ref{fig:JPiI_008}a.
}
\label{fig:JPiI_008}
\end{center}
\end{figure}

\begin{figure}[htb]
\begin{center}
\includegraphics*[height=9cm,width=7cm]{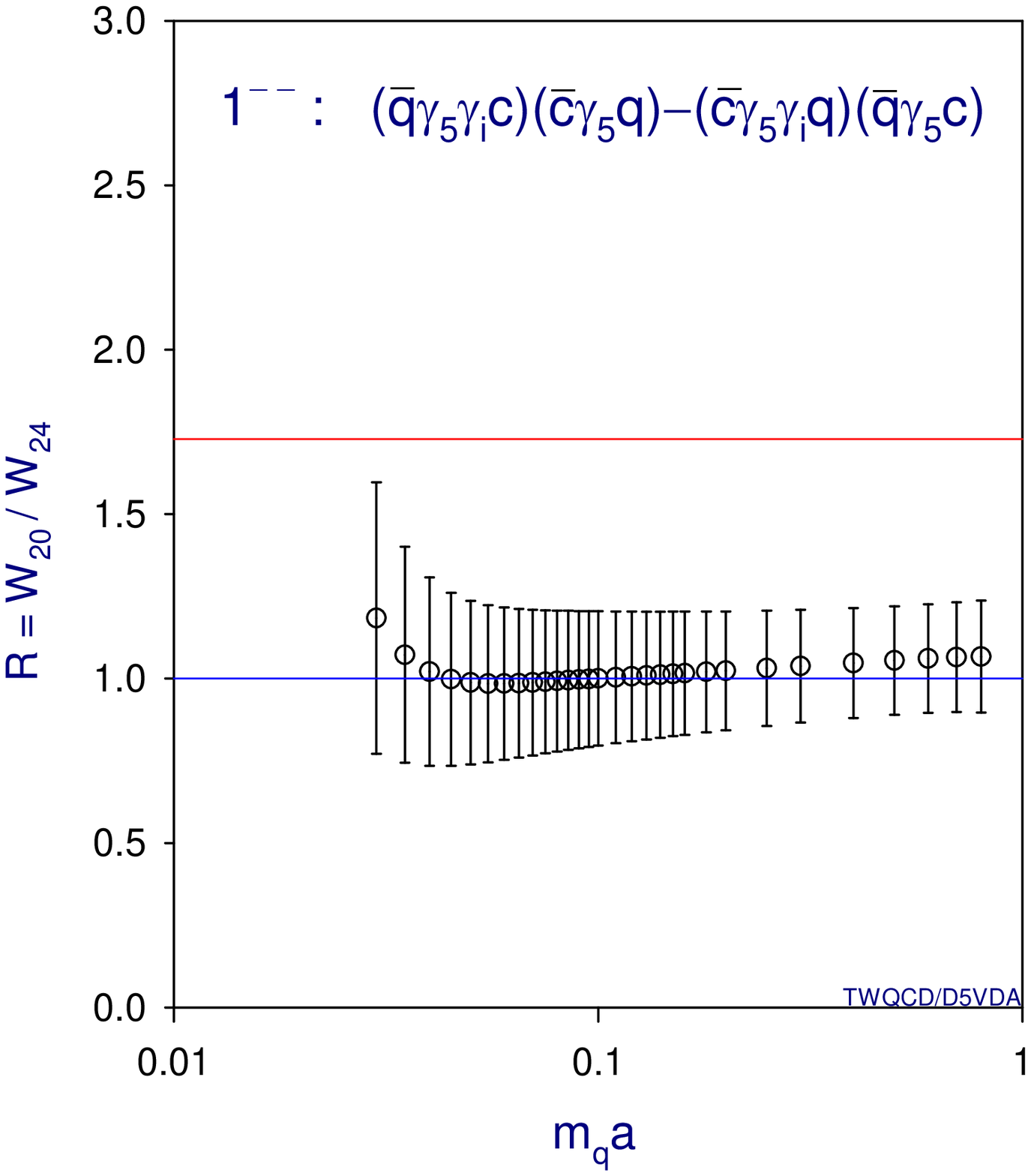}
\caption{
The ratio of spectral weights of the lowest-lying state
of the molecular operator $ M_3 $,
for $ 20^3 \times 40 $ and $ 24^3 \times 48 $ lattices at $ \beta = 6.1 $.
The upper-horizontal line $ R = (24/20)^3 = 1.728 $,
is the signature of 2-particle scattering state,
while the lower-horizontal line $ R = 1.0 $ is the signature
of a resonance.}
\label{fig:sw2024_D5VDA}
\end{center}
\end{figure}

\begin{figure}[htb]
\begin{center}
\includegraphics*[height=9cm,width=7cm]{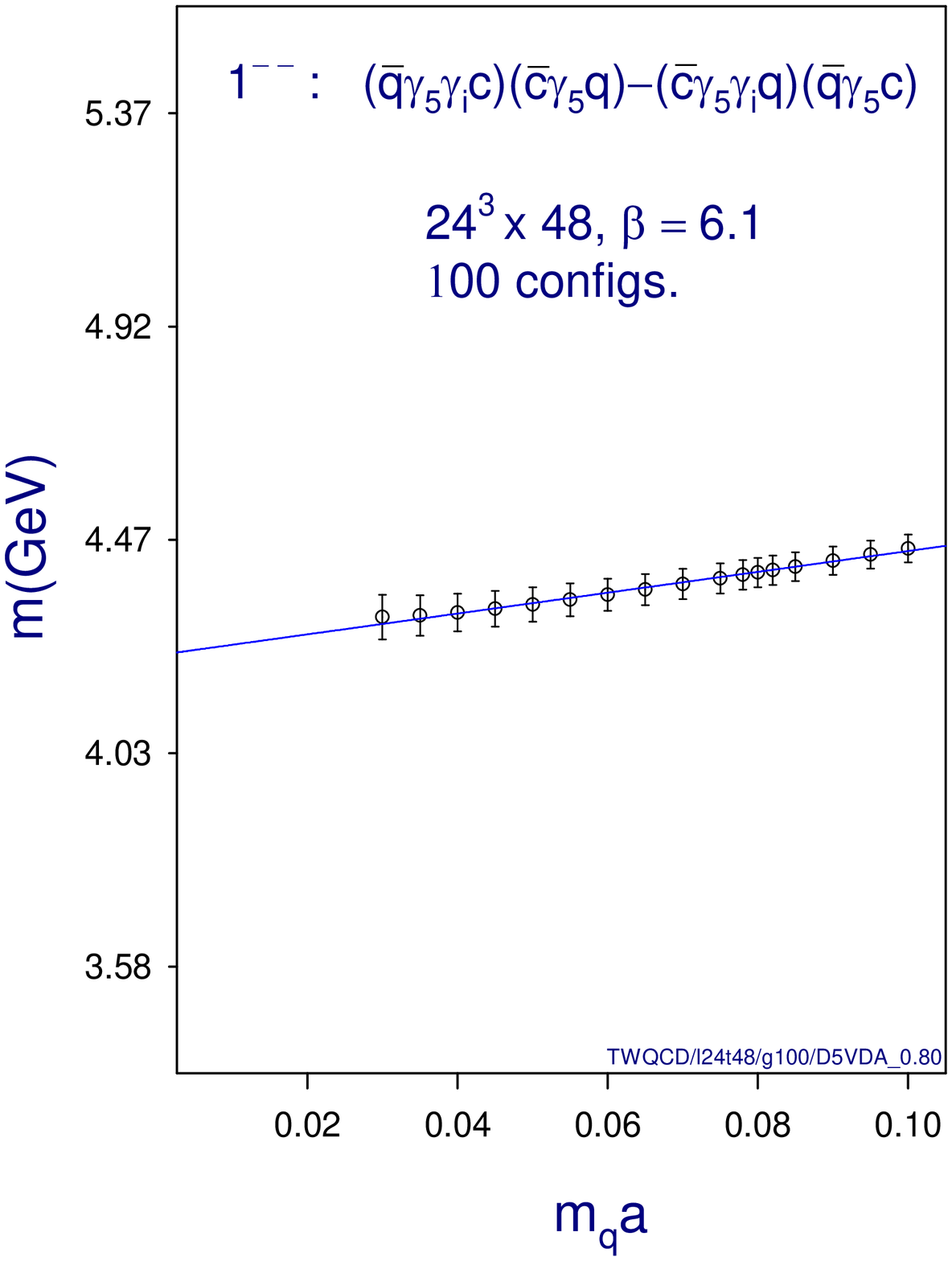}
\caption{
The mass of the lowest-lying state
of $ M_3 $ versus the quark mass $ m_q a $,
on the $ 24^3 \times 48 $ lattice at $ \beta = 6.1 $.
The solid line is the linear fit with $ \chi^2/d.o.f. = 0.25 $.}
\label{fig:mass_D5VDA}
\end{center}
\end{figure}

\begin{figure}[htb]
\begin{center}
\begin{tabular}{@{}cc@{}}
\includegraphics*[height=9cm,width=7cm]{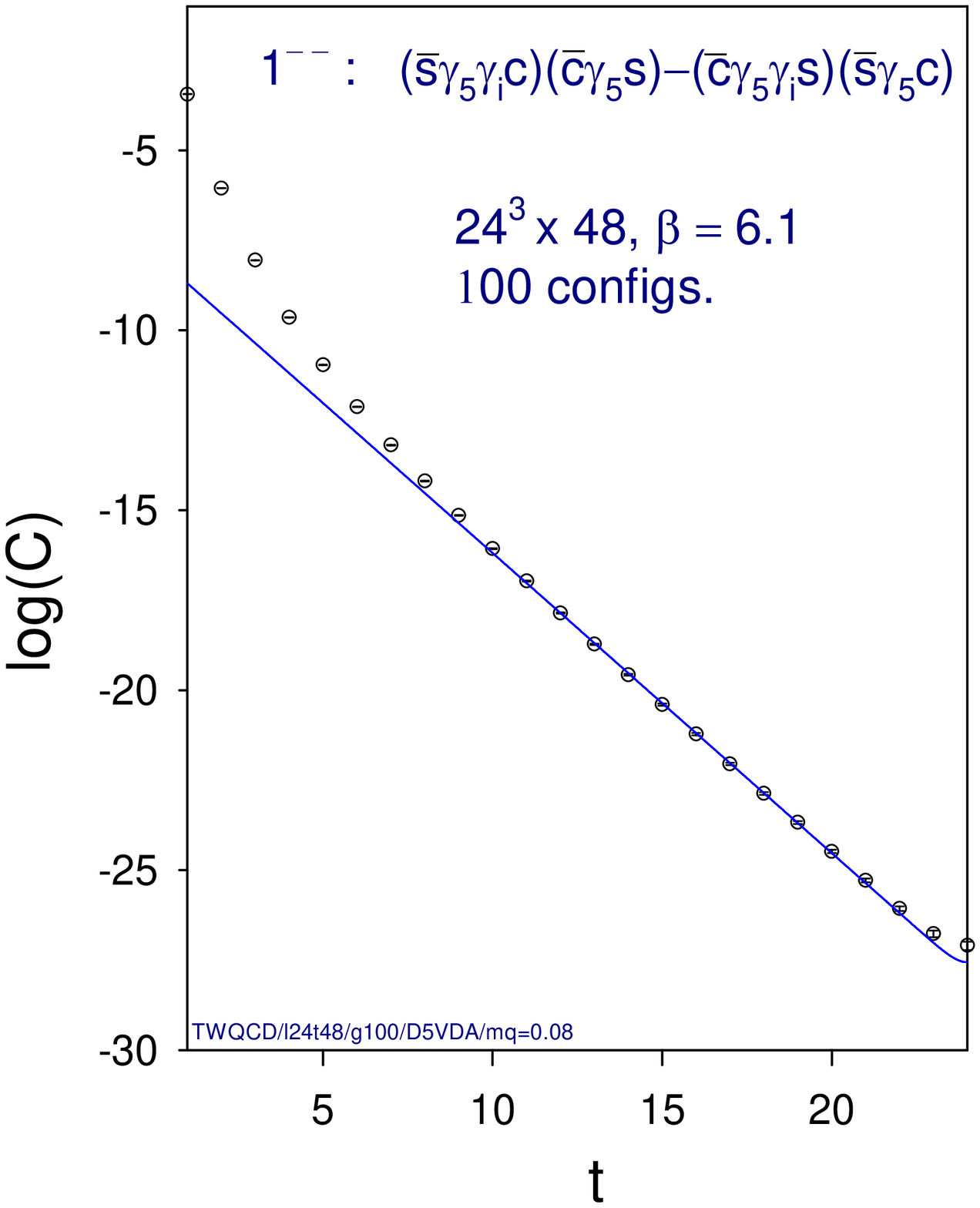}
&
\includegraphics*[height=9cm,width=7cm]{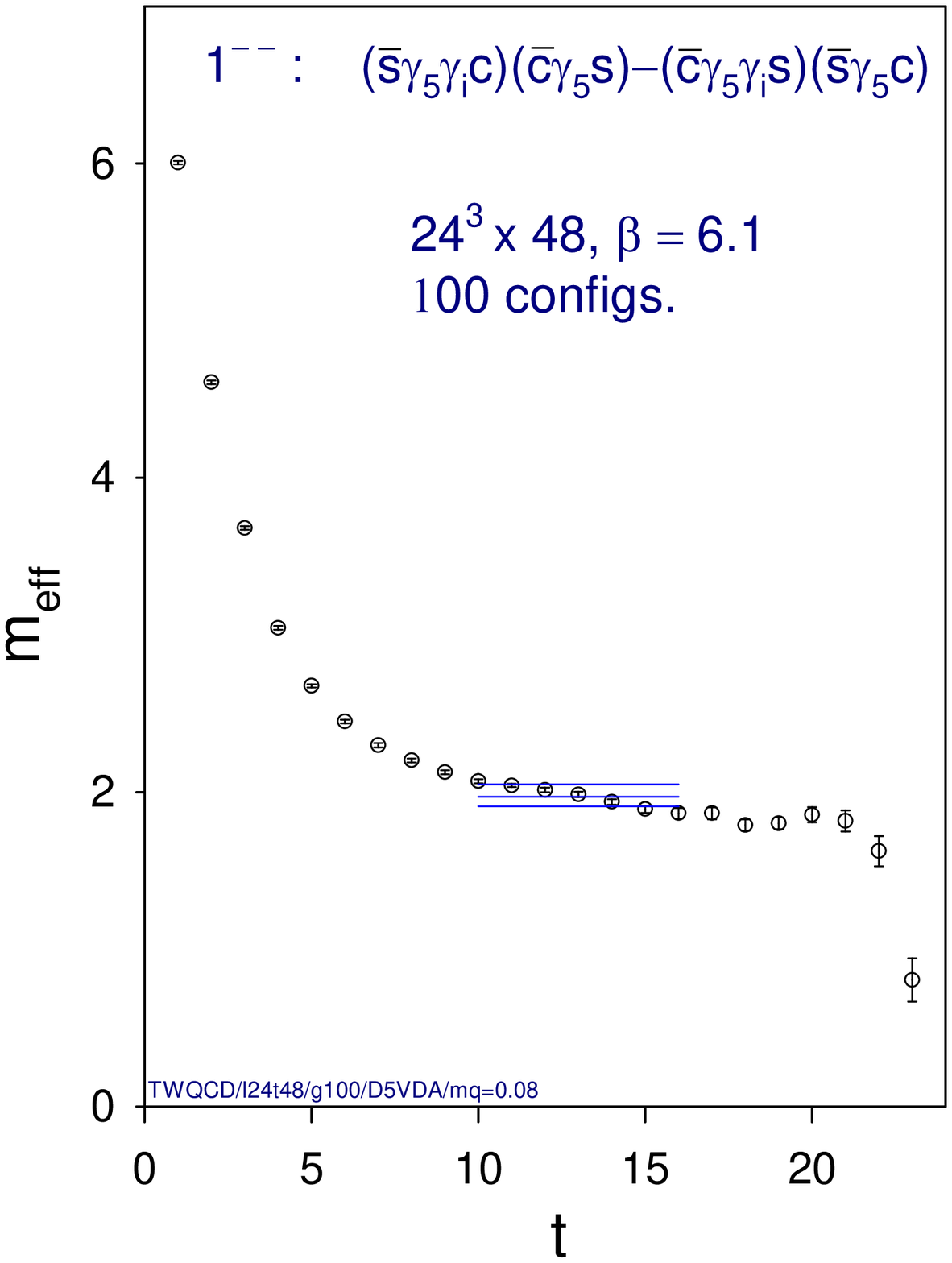}
\\ (a) & (b)
\end{tabular}
\caption{
(a) The time-correlation function $ C(t) $ of the lowest-lying state
of $ M_3 $ for $ m_q = m_s = 0.08 a^{-1} $,
on the $ 24^3 \times 48 $ lattice at $ \beta = 6.1 $.
The solid line is the hyperbolic-cosine fit for $ t \in [10,16] $
with $ \chi^2/d.o.f. = 1.07 $.
(b) The effective mass $ M_{eff}(t) = \ln [C(t)/C(t+1)]  $
of $ C(t) $ in Fig.\ \ref{fig:D5VDA_008}a.
}
\label{fig:D5VDA_008}
\end{center}
\end{figure}

The time-correlation function\footnote{
Here we have neglected the 
$ \c\cbar $ and $ \q\qbar $ annihilation diagrams 
such that $ C(t) $ does not overlap with any conventional meson  
($ \c\cbar $ or $ \q\qbar $) states.}  
\BAN
C_{M}(t)=\sum_{\vec{x}} \left< M(\vec{x},t) M^\dagger(\vec{0},0) \right>
\EAN
is measured for each gauge configuration, and its average over all 
gauge configurations is fitted to the usual formula 
\BAN
\frac{Z}{2 m a } [ e^{-m a t} + e^{-m a (T-t)} ]
\EAN
to extract the mass $ m a $ of the lowest-lying state 
and its spectral weight
\BAN
W = \frac{Z}{2 m a } \ .
\EAN
Theoretically, if this state is a genuine resonance, then its mass $ m a $  
and spectral weight $ W $ should be almost constant for 
any lattices with the same lattice spacing. On the 
other hand, if it is a 2-particle scattering state, then its mass 
$ m a $ is sensitive to the lattice volume, and its spectral
weight is inversely proportional to the spatial volume for lattices
with the same lattice spacing. In the following, we shall
use the ratio of the spectral weights on two spatial volumes
$ 20^3 $ and $ 24^3 $ with the same lattice spacing ($\beta = 6.1 $) 
to discriminate whether any hadronic state under investigation 
is a resonance or not.

In Fig. \ref{fig:sw2024_DVDIS}, the ratio ($ R=W_{20}/W_{24} $) 
of spectral weights of the lowest-lying state extracted from 
the time-correlation function of $ M_1 $ on the 
$ 20^3 \times 40 $ and $ 24^3 \times 48 $ lattices is plotted 
versus the quark mass $ m_q a \in [0.03, 0.8] $. 
Evidently, $ R \simeq 1.0 $ for $ m_q a > 0.05 $, 
which implies that there exist $ 1^{--} $ resonances with quark 
contents $ (\c\c\cbar\cbar) $ and $ (\c\s\cbar\sbar) $.  
On the other hand, as $ m_q \to m_{u} $, $ R $ begins to 
deviate from 1.0 with large error.  
We suspect that this might be due to quenching artifacts as 
$ m_q \to m_{u} $, mostly from the scalar meson component 
$ (\cbar \q) $ or $ (\qbar \c) $ in the molecular operator $ M_1 $.   
Thus $ R $ could be consistent with 1.0 
if one incorporates internal quark loops, with higher statistics 
and larger volumes. If this is the case, 
a resonance might also exist for $ \q = \u $. 
In the following, we assume this is the case, and chirally extrapolate 
the mass of the molecule $ M_1 $ to the limit $ m_q \to m_{u} $.

In Fig. \ref{fig:mass_DVDIS}, the mass of the lowest-lying state 
extracted from the molecular operator $ M_1 $ is plotted versus $ m_q a $, 
which can be fitted by the linear function $ m = c_0 + c_1 m_q $ 
with $ \chi^2/d.o.f. = 0.34 $. 
In the limit $ m_q \to m_{u} $, it gives $ m = 4350(69) $ MeV 
which is compatible with the mass of $ Y(4260) $.

For $ m_q = m_s  $, and $ m_q = m_c $,  
the time-correlation functions 
and effective masses of $ M_1 $ are plotted in 
Fig. \ref{fig:DVDIS_008}  
and Fig. \ref{fig:DVDIS_080} respectively.
The masses of the lowest-lying states are:
$m[(\sbar\gi\c)(\cbar\s)+(\cbar\gi\s)(\sbar\c)]=4546(30)$ MeV with 
$ \chi^2/d.o.f. = 0.97 $, 
and $m[(\cbar\gi\c)(\cbar\c)]=6411(25)$ MeV with   
$ \chi^2/d.o.f. = 0.61 $.

Next we turn to the molecular operator $ M_2 $. Obviously, it must 
suffer severely from the quenched artifacts as $ m_q \to m_{u} $, 
due to the presence of the scalar meson operator $ (\qbar \q) $.
Thus, we observe that in the limit $ m_q \to m_{u} $, 
the time-correlation function becomes very noisy, 
and we could not obtain any reliable results of its
masses and spectral weights. However, at $ m_q = m_s = 0.08 a^{-1} $, 
we can still extract its mass and spectral weight. 
For the $ 24^3 \times 48 $ lattice (Fig. \ref{fig:JPiI_008}), 
the mass and spectral weight 
are: $ m = 4581(96) $ MeV, and $ W = 0.71(41) \times 10^{-7} $; 
while for the $ 20^3 \times 40 $
lattice, $ m = 4637(156) $ MeV, and $ W = 0.92(80) \times 10^{-7}$.  
Thus the molecular operator $ (\cbar \gamma_i \c) (\sbar \s) $ 
seems to detect a resonance with mass $ 4581(96) $ MeV 
(Fig. \ref{fig:JPiI_008}), which is compatible to that obtained from 
$ M_1 $ with $ m_q = m_s $ (Fig. \ref{fig:DVDIS_008}).

Next we consider the molecular operator $ M_3 $. 
It is expected to suffer quenched artifacts less than those of 
$ M_1 $ and $ M_2 $, since it is composed of a pseudoscalar meson 
operator $ (\cbar \gamma_5 q) $ times a pseudovector meson operator
$ (\qbar \gamma_5 \gamma_i c) $, without any scalar meson operators 
like $ (\qbar \q) $ or $ (\cbar \q) $. 
In Fig. \ref{fig:sw2024_D5VDA}, the ratio ($ R = W_{20}/W_{24} $) 
of spectral weights of the lowest-lying state extracted from 
the time-correlation function of $ M_3 $ on the 
$ 20^3 \times 40 $ and $ 24^3 \times 48 $ lattices is plotted 
versus the quark mass $ m_q a \in [0.03, 0.8] $. 
Evidently, $ R \simeq 1.0 $ for the entire range of quark masses,   
which implies that there exist resonances with quark contents
$ (\c\c\cbar\cbar) $, $ (\c\s\cbar\sbar) $, and $ (\c\u\cbar\ubar) $,     
even though $ R $ slightly deviates from one at small quark masses.

In Fig. \ref{fig:mass_D5VDA}, the mass of the lowest-lying state 
extracted from the molecular operator $ M_3 $ is plotted versus 
$ m_q a $, which can be fitted by the linear function 
$ m = c_0 + c_1 m_q $ with $ \chi^2/d.o.f. = 0.25 $. 
In the limit $ m_q \to m_{u} $, it gives $ m = 4238(31) $ MeV, 
which is in good agreement with the mass of $ Y(4260) $.

For $ m_q = m_s = 0.08 a^{-1} $, the time-correlation function 
and effective mass of $ M_3 $ are plotted in Fig. \ref{fig:D5VDA_008}.  
The mass of the lowest-lying state is $ 4405(31)$ MeV,  
which is compatible to those obtained from 
$ M_1 $ and $ M_2 $ with $ m_q = m_s $.

\section{The Diquark-Antidiquark Operator}

\begin{figure}[htb]
\begin{center}
\includegraphics*[height=9cm,width=7cm]{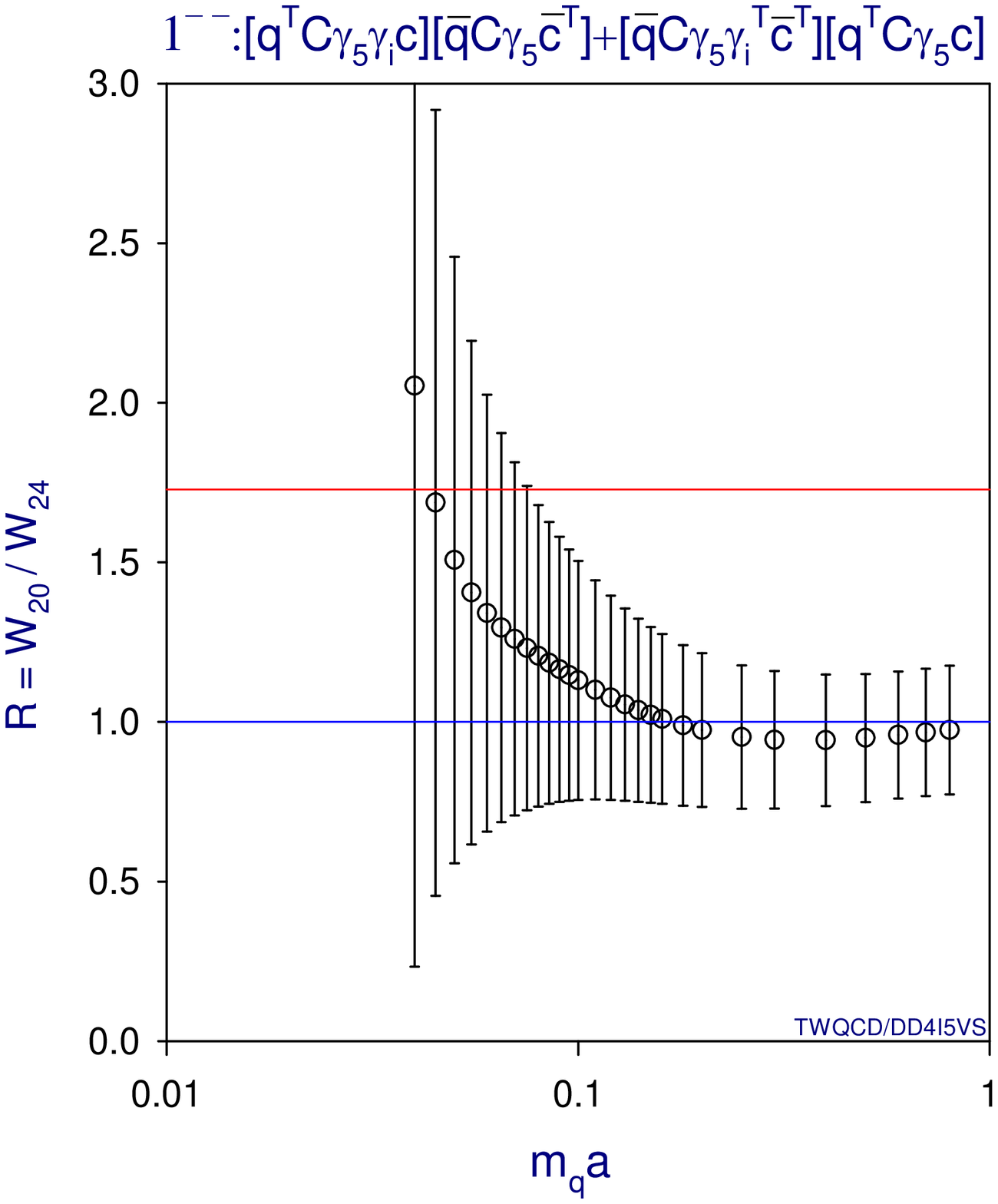}
\caption{
The ratio of spectral weights of the lowest-lying state
of diquark-antidiquark operator $ Y_4 $, 
for $ 20^3 \times 40 $ and $ 24^3 \times 48 $ lattices at $ \beta = 6.1 $.
The upper-horizontal line $ R = (24/20)^3 = 1.728 $,
is the signature of 2-particle scattering state,
while the lower-horizontal line $ R = 1.0 $ is the signature
of a resonance.}
\label{fig:sw2024_DD4I5VS}
\end{center}
\end{figure}

\begin{figure}[htb]
\begin{center}
\includegraphics*[height=9cm,width=7cm]{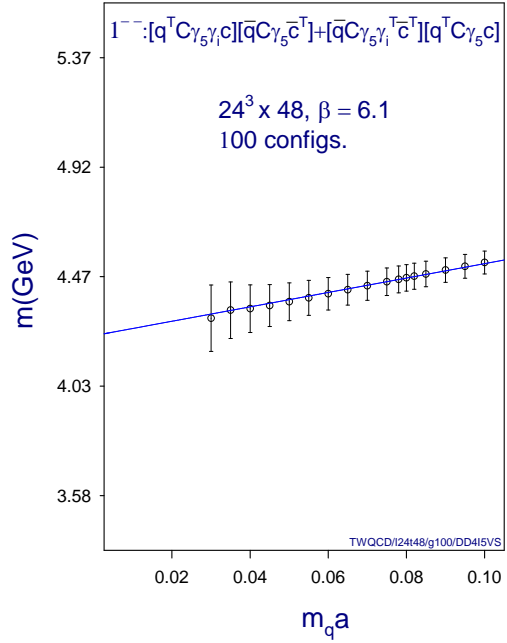}
\caption{
The mass of the lowest-lying state of the diquark-antidiquark operator
$ Y_4 $ versus the quark mass $ m_q a $,
on the $ 24^3 \times 48 $ lattice at $ \beta = 6.1 $.
The solid line is the linear fit with $ \chi^2/d.o.f. = 0.49 $.}
\label{fig:mass_DD4I5VS}
\end{center}
\end{figure}

\begin{figure}[htb]
\begin{center}
\begin{tabular}{@{}cc@{}}
\includegraphics*[height=9cm,width=7cm]{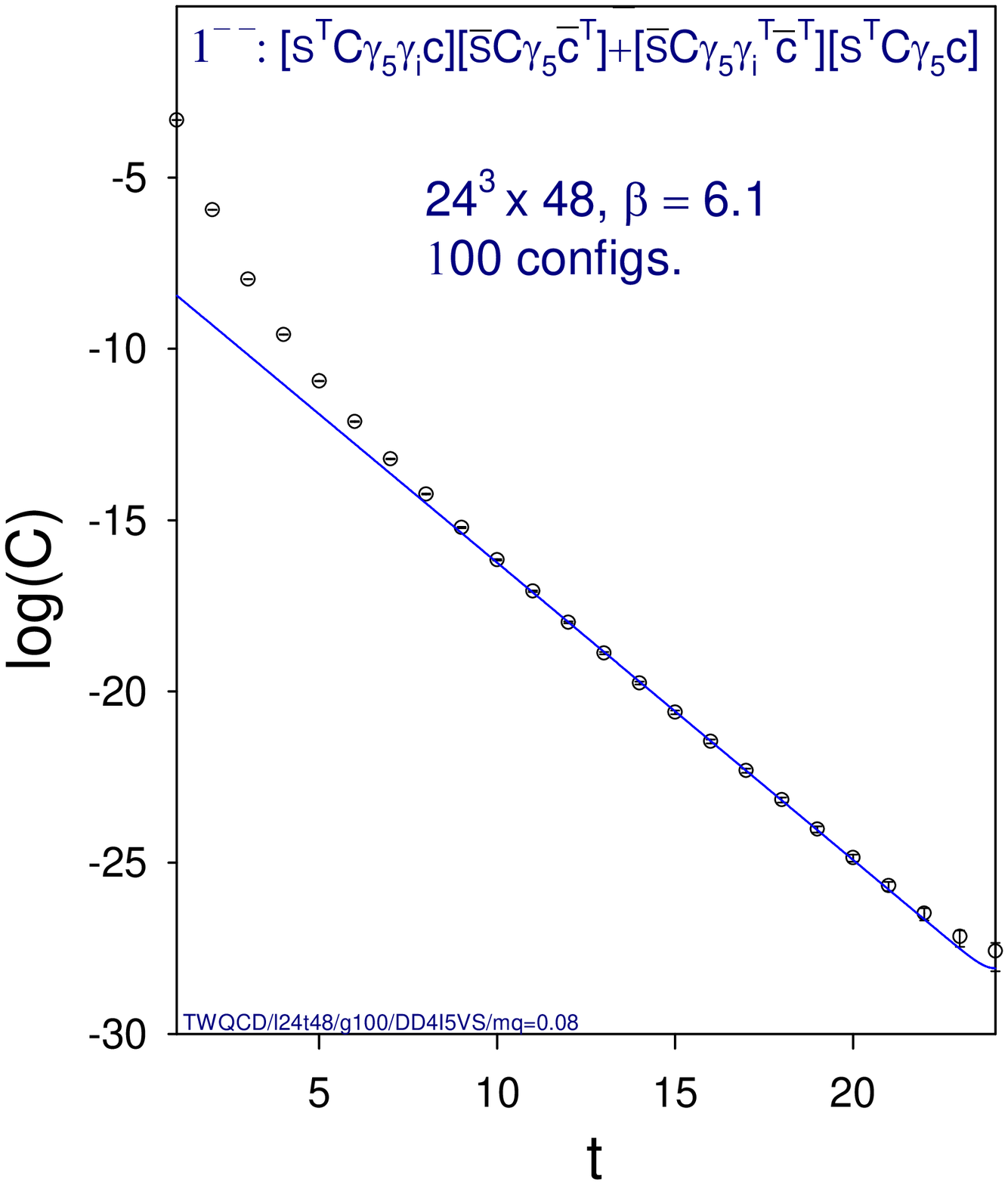}
&
\includegraphics*[height=9cm,width=7cm]{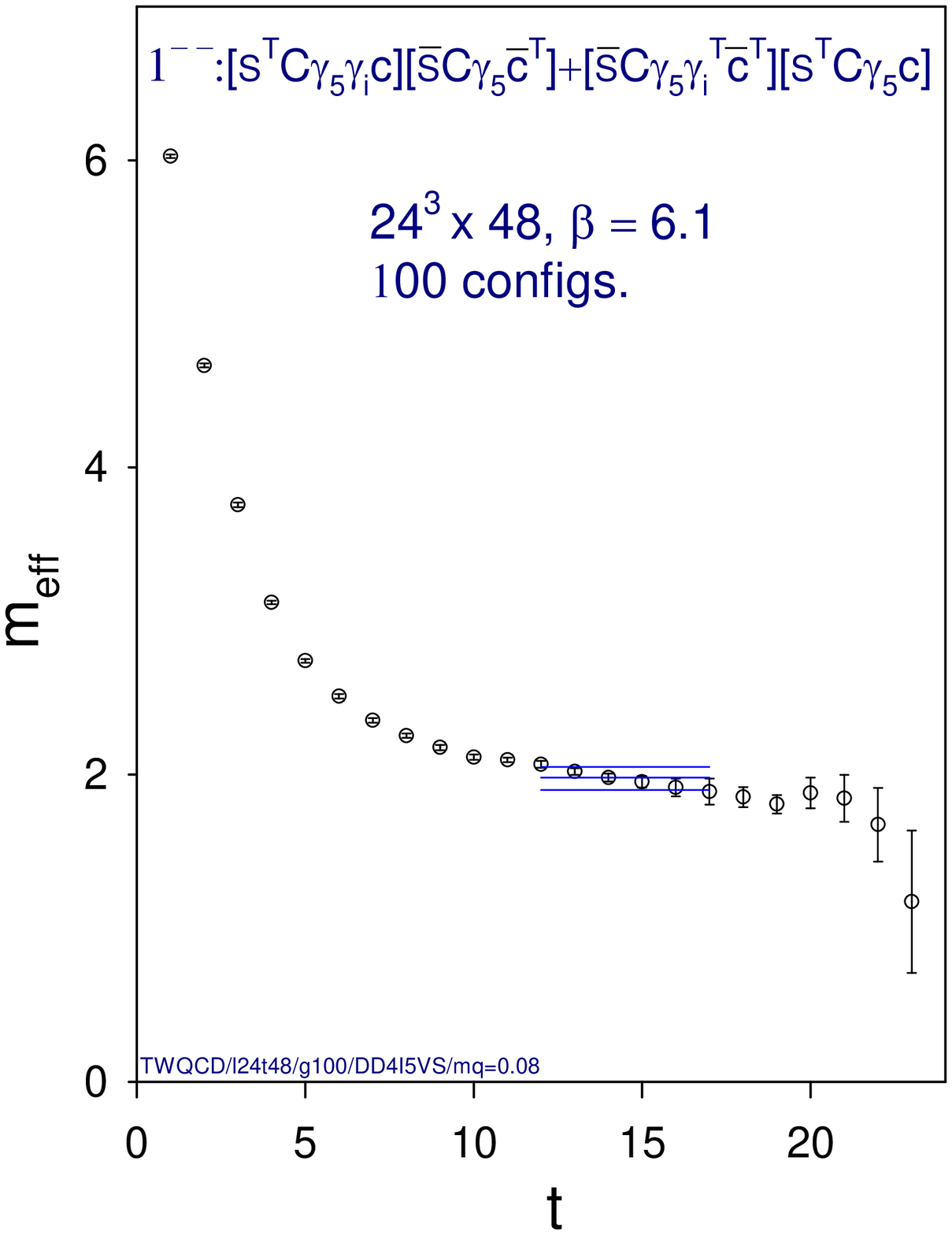}
\\ (a) & (b)
\end{tabular}
\caption{
(a) The time-correlation function $ C(t) $ of the lowest-lying state
of $ Y_4 $ for $ m_q = m_s = 0.08 a^{-1} $,
on the $ 24^3 \times 48 $ lattice at $ \beta = 6.1 $.
The solid line is the hyperbolic-cosine fit for $ t \in [12,17] $
with $ \chi^2/d.o.f. = 1.03 $.
(b) The effective mass $ M_{eff}(t) = \ln [C(t)/C(t+1)]  $
of $ C(t) $ in Fig.\ \ref{fig:DD4I5VS_008}a.
}
\label{fig:DD4I5VS_008}
\end{center}
\end{figure}

\begin{figure}[htb]
\begin{center}
\begin{tabular}{@{}cc@{}}
\includegraphics*[height=9cm,width=7cm]{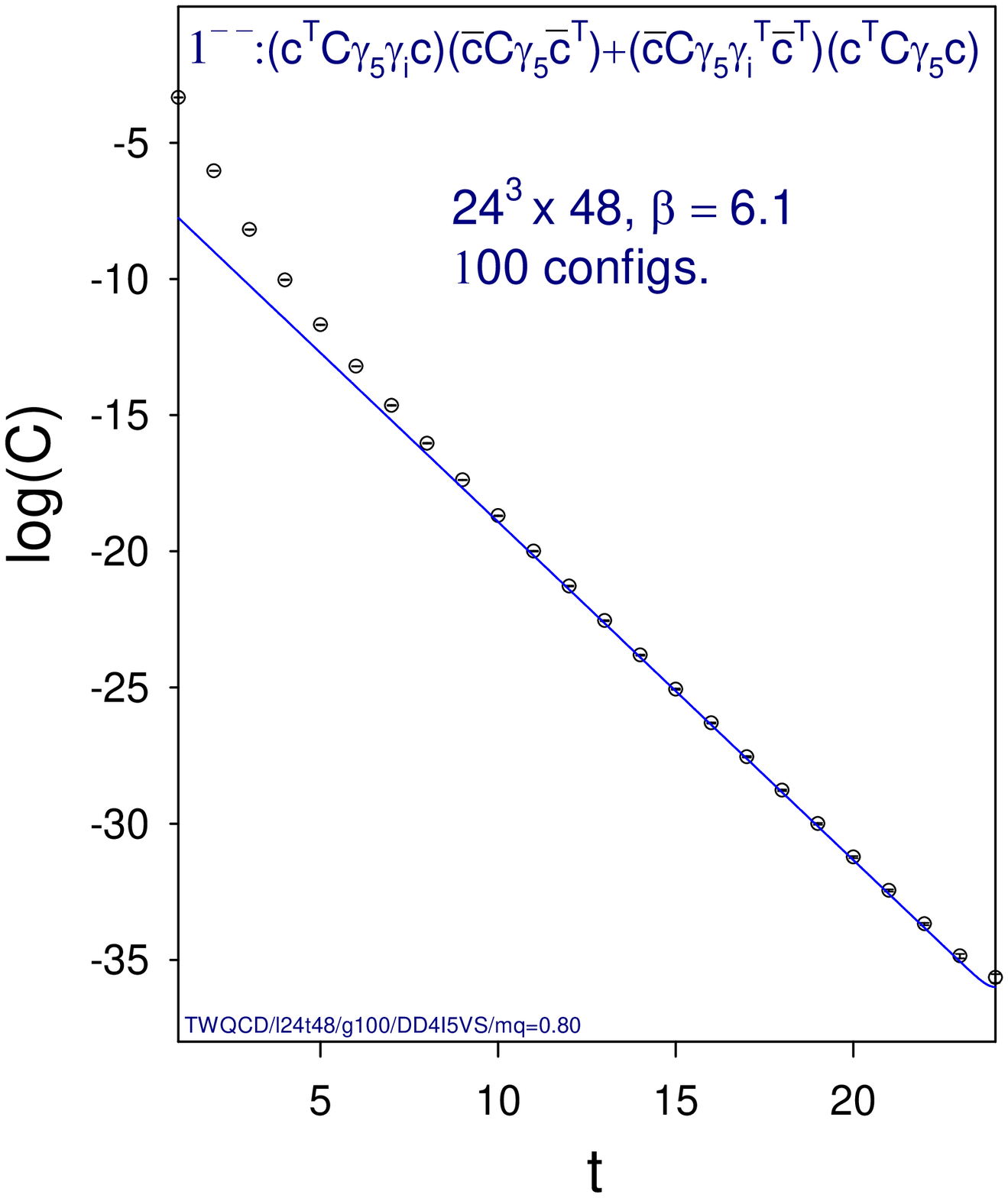}
&
\includegraphics*[height=9cm,width=7cm]{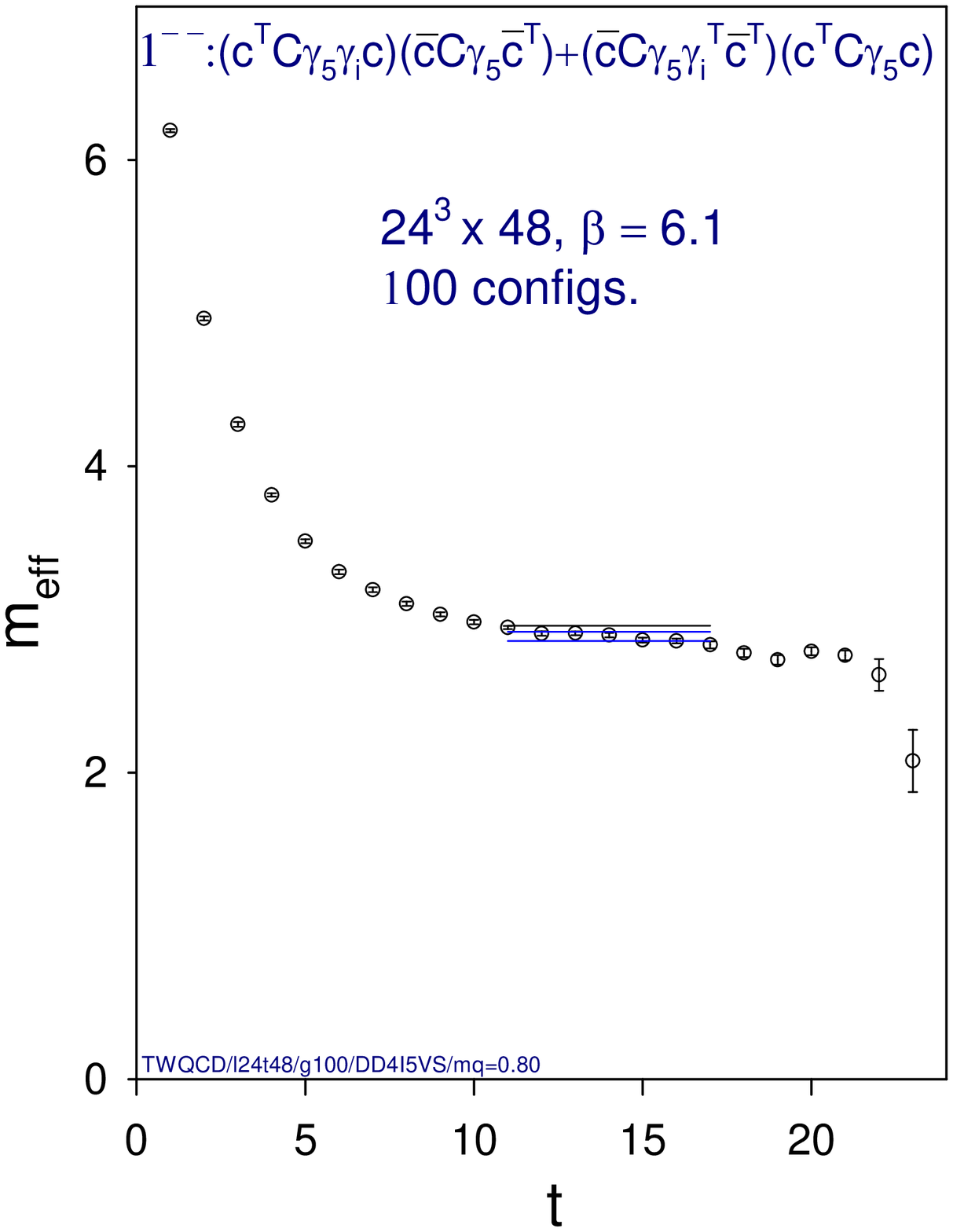}
\\ (a) & (b)
\end{tabular}
\caption{
(a) The time-correlation function $ C(t) $ of the lowest-lying state
of $ Y_4 $ for $ m_q = m_c = 0.80 a^{-1} $,
on the $ 24^3 \times 48 $ lattice at $ \beta = 6.1 $.
The solid line is the hyperbolic-cosine fit for $ t \in [11,17] $
with $ \chi^2/d.o.f. = 0.92 $.
(b) The effective mass $ M_{eff}(t) = \ln [C(t)/C(t+1)]  $
of $ C(t) $ in Fig.\ \ref{fig:DD4I5VS_080}a.
}
\label{fig:DD4I5VS_080}
\end{center}
\end{figure}

We construct the diquark-antidiquark operator with $ J^{PC} = 1^{--} $ as 
\bea
\label{eq:DD4I5VS}
Y_4(x)
\EQ \frac{1}{\sqrt{2}} \left\{ 
   [\q^T C \gamma_5 \gi \c ]_{xa}[\qbar C \gamma_5 \cbar^T]_{xa} 
  +[\qbar C \gamma_5 \gi^T \cbar^T]_{xa}[\q^T C \gamma_5 \c]_{xa} \right\} 
\eea
where $ C $ is the charge conjugation operator satisfying
$ C \gamma_\mu C^{-1} = -\gamma_\mu^T $ and
$ (C \gamma_5)^T=-C\gamma_5 $. Here the 
diquark operator $ [\q^T \Gamma \Q]_{xa} $ for any Dirac matrix 
$ \Gamma $ is defined as
\bea
\label{eq:diquark}
[{\q}^T \Gamma {\Q} ]_{xa} \equiv \epsilon_{abc} (
 {\q}_{x\alpha b} \Gamma_{\alpha\beta} {\Q}_{x\beta c}
-{\Q}_{x\alpha b} \Gamma_{\alpha\beta} {\q}_{x\beta c} )
\eea
where 
$ x $, $ \{ a,b,c \} $ and $ \{ \alpha, \beta \} $
denote the lattice site, color, and Dirac indices respectively,  
and $ \epsilon_{abc} $ is the completely antisymmetric tensor. 
Thus the diquark (\ref{eq:diquark}) transforms like color anti-triplet.  
For $ \Gamma = C \gamma_5 $, it transforms like $ J^P = 0^{+} $,  
while for $ \Gamma = C \gamma_5 \gamma_i $ ($i=1,2,3 $), it 
transforms like $ 1^{-} $. 
In the limit $ \q = \Q $, the diquark operator is replaced by 
$
({\q}^T \Gamma {\q} )_{xa} \equiv \epsilon_{abc} 
{\q}_{x\alpha b} \Gamma_{\alpha\beta} {\q}_{x\beta c}
$.

In Fig. \ref{fig:sw2024_DD4I5VS}, the ratio ($ R=W_{20}/W_{24} $)
of spectral weights of the lowest-lying state extracted from
the time-correlation function of $ Y_4 $ on the $ 20^3 \times 40 $
and $ 24^3 \times 48 $ lattices is plotted
versus the quark mass $ m_q a \in [0.03, 0.8] $.
Evidently, $ R \simeq 1.0 $ for $ m_q a > 0.05 $,
in particular, for $ m_q = m_s = 0.08 a^{-1} $,
and $ m_q = m_c = 0.8 a^{-1} $. This implies that 
that there exist $ 1^{--} $ resonances with quark 
contents $ (\c\c\cbar\cbar) $ and $ (\c\s\cbar\sbar) $.  
On the other hand,
as $ m_q \to m_{u} $, $ R $ begins to
deviate from 1.0 with large error.
This seems to suggest that the diquark-antidiquark operator $ Y_4 $
has little overlap with the resonance detected by the molecular operator 
$ M_3 $ as $ m_q \to m_u $. 
However, it is unclear whether $ R \simeq 1 $ if one incorporates
internal quark loops, and with larger volumes and higher statistics. 
In the following, we assume this is the case, 
and obtain its mass by chiral extrapolation.

In Fig. \ref{fig:mass_DD4I5VS}, the mass of the lowest-lying state
of the diquark-antidiquark operator $ Y_4 $
is plotted versus $ m_q a $, which can be fitted by the linear function 
$ m = c_0 + c_1 m_q $ with $ \chi^2/d.o.f. = 0.49 $. 
In the limit $ m_q \to m_{u} $, it gives
$ m = 4267(68) $ MeV, which is in good agreement
with the mass of $ Y(4260) $.

For $ m_q = m_s = 0.08 a^{-1} $, and $ m_q = m_c = 0.80 a^{-1} $, 
the time-correlation functions and effective masses of the 
diquark-antidiquark operator are plotted in Fig. \ref{fig:DD4I5VS_008},  
and Fig. \ref{fig:DD4I5VS_080} respectively. 
The masses of the lowest-lying states are:
$ m\{
     [\s^T C \gamma_5 \gi \c ][\sbar C \gamma_5 \cbar^T] 
    +[\sbar C \gamma_5 \gi^T \cbar^T][\s^T C \gamma_5 \c]
\} = 4449(40) $ MeV
with $ \chi^2/d.o.f. = 1.03 $, 
and $ m\{
   (\c^T C \gamma_5 \gi \c)(\cbar C \gamma_5 \cbar^T) 
  +(\cbar C \gamma_5 \gi^T \cbar^T)(\c^T C \gamma_5 \c) \}
) = 6420(29) $ MeV
with $ \chi^2/d.o.f. = 0.92 $.

Besides $ Y_4 $, we have also constructed other diquark-antidiquark operators, 
e.g., 
\BAN
\label{eq:DD4VIS}
\frac{1}{\sqrt{2}} \left\{ 
   [\q^T C \gi \c ]_{xa}[\qbar C \cbar^T]_{xa} 
  +[\qbar C \gi^T \cbar^T]_{xa}[\q^T C \c]_{xa} \right\} 
\EAN
to see whether they have good overlap with any resonances as $ m_q \to m_u $. 
However, it turns out that they all behave similar to $ Y_4 $.

\section{Summary and Discussions}

\begin{table}
\begin{center}
\begin{tabular}{c|c|c}
Operator & Mass (MeV) & Resonance/Scattering \\
\hline
\hline
$\epsilon_{ijk}\cbar\gamma_5 F_{jk}\c $ & 4501(178)(215) & Resonance \\
\hline
$\frac{1}{\sqrt{2}}[ (\ubar \gamma_i \c)(\cbar \u)
                   +(\cbar \gamma_i \u)(\ubar \c) ] $
        &   4350(69)(88) &   Resonance ? \\
$ \frac{1}{\sqrt{2}}[ (\sbar \gamma_i \c)(\cbar \s)
                   +(\cbar \gamma_i \s)(\sbar \c) ]  $
        &   4546(30)(61) &    Resonance       \\
$ \frac{1}{\sqrt{2}}[ (\ubar \gamma_5 \gamma_i \c)(\cbar \gamma_5 \u)
                   -(\cbar \gamma_5 \gamma_i \u)(\ubar \gamma_5 \c) ] $
        &   4238(31)(57)  &   Resonance  \\
$ \frac{1}{\sqrt{2}}[ (\sbar \gamma_5 \gamma_i \c)(\cbar \gamma_5 \s)
                   -(\cbar \gamma_5 \gamma_i \s)(\sbar \gamma_5 \c) ] $
        &  4405(31)(44)  & Resonance  \\
$ (\cbar \gamma_i \c) (\sbar \s) $   &   4581(96)(115) & Resonance \\
$ (\cbar \gamma_i \c) (\cbar \c) $   &   6411(25)(43) & Resonance \\
\hline
$ \frac{1}{\sqrt{2}}\left\{[\u^T C\gamma_5\gamma_i \c][\ubar C\gamma_5\cbar^T]
  +[\u^T C \gamma_5 \c][\ubar C\gamma_5\gamma_i^T \cbar^T] \right\} $      &
4267(68)(83)  & Resonance ? \\
$ \frac{1}{\sqrt{2}} \left\{[\s^T C\gamma_5\gamma_i\c][\sbar C\gamma_5\cbar^T]
  +[\s^T C \gamma_5 \c][\sbar C\gamma_5\gamma_i^T \cbar^T] \right\} $      &
4449(40)(55) & Resonance  \\
$ \frac{1}{\sqrt{2}}\left\{(\c^T C\gamma_5\gamma_i\c)(\cbar C\gamma_5 \cbar^T)
  +(\c^T C \gamma_5 \c)(\cbar C\gamma_5\gamma_i^T \cbar^T) \right\} $      &
6420(29)(32) & Resonance  \\
\hline
\hline
\end{tabular}
\caption{Mass spectra of hybrid charmonium, molecules, 
and diquark-antidiquark operators with $ J^{PC} = 1^{--} $.}
\label{tab:mass_summary}
\end{center}
\end{table}

In this paper, we have investigated the mass spectra of 
several interpolating operators 
(i.e., the hybrid charmonium $ \c \cbar g $, 
the molecular operators $ M_1 $, $ M_2 $, and $ M_3 $, 
and the diquark-antidiquark operator $ Y_4 $) with 
$ J^{PC} = 1^{--} $, in quenched lattice QCD with exact chiral symmetry. 
Our results are summarized in Table \ref{tab:mass_summary}, 
where in each case, the first error is statistical, and 
the second one is our estimate of combined systematic uncertainty
including those coming from: 
(i) possible plateaus (fit ranges) with $ \chi^2/d.o.f. < 1 $; 
(ii) the uncertainties in the strange quark mass and the charm quark mass; 
(iii) chiral extrapolation (for the entries containing u/d quarks); and
(iv) finite size effects (by comparing results of two lattice sizes).
Note that we cannot estimate the discretization error  
since we have been working with one lattice spacing. 
Even though lattice QCD with exact chiral symmetry does not have 
$ O(a) $ and $ O(ma) $ lattice artifacts, the $ O(m^2 a^2) $ effect 
might turn out to be not negligible for $ m_c a = 0.8 $. 
 
For the hybrid charmonium ($\epsilon_{ijk}\cbar\gamma_5 F_{jk}\c $), 
the mass of the lowest-lying state 
only agrees marginally with the mass of $ Y(4260) $.  
Thus it is unlikely to be identified with $ Y(4260) $, 
even though we cannot rule out such a possibility. 
Nevertheless, we hope to pin down this problem with a more precise 
determination of the spectrum of the $ 1^{--} $ hybrid charmonium 
in a future publication.

Evidently, the molecular operator 
$ M_3 \sim \{ (\qbar \gamma_5 \gamma_i \c)(\cbar \gamma_5 \q)
    -(\cbar \gamma_5 \gamma_i \q)(\qbar \gamma_5 \c) \} $
detects a resonance ($ J^{PC} = 1^{--} $) with mass $ 4238(31)(57) $ MeV  
in the limit $ m_q \to m_u $, which is naturally identified 
with $ Y(4260) $. 
This seems to suggest that $ Y(4260) $ is indeed in the spectrum of QCD, 
with quark content ($\c\u\cbar\ubar$). 
Note that we have not studied 
the excited states of $ c \cbar $, thus we could not rule out 
the possibility that $ Y(4260) $ might turn out to be one of 
the excited states of $ \c\cbar $, e.g., $ \psi(4^3 S_1) $
or $ \psi(3^3 D_1) $, even though this is very unlikely 
in view of the widely accepted experimental and theoretical 
spectrum of $ c \cbar $.

For the molecular operator
$ M_1 \sim \{(\qbar \gamma_i \c)(\cbar \q)
                   +(\cbar \gamma_i \q)(\ubar \c) \} $, 
and the diquark-antidiquark operator   
$ Y_4 \sim \{[\q^T C\gamma_5\gamma_i \c][\qbar C\gamma_5\cbar^T]
   +[\q^T C\gamma_5 \c][\qbar C\gamma_5\gamma_i^T \cbar^T] \} $, 
they also detect states with masses $ 4350(69)(88) $ MeV, and  
$ 4267(68)(83) $ MeV respectively, in the limit $ m_q \to m_u $.
We suspect that they might be the same resonance captured by 
the molecular operator $ M_3 $.    
However, we are not sure that these states are resonances since 
the ratio of spectral weights ($ R = W_{20} / W_{24} $) 
for two different lattice volumes with the same lattice spacing 
deviates from one (the criterion for a resonance) 
with large errors as $ m_q \to m_u $. 
It is plausible that such a deviation is due to the 
quenched artifacts which can be evaded if one incorporates
internal quark loops.

Now, in the quenched approximation, our results  
suggest that $ Y(4260) $ has a better overlap with the molecular 
operator $ M_3 $ than any other operators discussed in this paper. 
Whether this implies that $ Y(4260) $ behaves more likely as a 
$ D_1 \bar D $ molecule than other molecules or diquark-antidiquark 
meson is subjected to further investigations, especially those 
incorporating dynamical quarks.

Finally, all molecular and diquark-antidiquark operators with 
quark fields ($\c\s\cbar\sbar$) detect a resonance around 
$ 4450 \pm 100 $ MeV, and those with ($\c\c\cbar\cbar$) 
detect a resonance around $ 6400 \pm 50 $ MeV.  
These serve as predictions of lattice QCD with exact chiral symmetry.

\bigskip
\bigskip

\flushpar
{\bf Acknowledgement}

\noindent

\bigskip

This work was supported in part by the National Science Council,
Republic of China, under the Grant No. NSC94-2112-M002-016 (T.W.C.),  
and Grant No. NSC94-2119-M239-001 (T.H.H.), and by 
National Center for High Performance Computation at Hsinchu.
T.W.C. would like to thank the kind hospitality   
of Yukawa Institute for Theoretical Physics at Kyoto University, 
where the final version of this paper was prepared during 
the YITP workshop "Actions and symmetries in lattice gauge theory"
(YITP-W-05-25).

\bigskip
\bigskip

\vfill\eject

\end{document}